\documentclass[12pt,preprint]{aastex}

\slugcomment{}

\shorttitle{IC2395}
\shortauthors{Balog et al.}

\begin{document}

\title{Protoplanetary and Transitional Disks in the Open Stellar Cluster IC\,2395}

\author{Zoltan Balog\altaffilmark{1}, Nick Siegler\altaffilmark{2}, G. H. Rieke\altaffilmark{3}, L. L. Kiss\altaffilmark{4}, James Muzerolle\altaffilmark{5}, R. A. Gutermuth\altaffilmark{6}, Cameron P. M. Bell\altaffilmark{7}, J. Vink\'o\altaffilmark{8}, K. Y. L. Su\altaffilmark{3}, E. T. Young\altaffilmark{9}, Andr\'as G\'asp\'ar\altaffilmark{3}}

\altaffiltext{1}{Max Planck Institute for Astronomy, Heidelberg, D-69117, Germany}
\altaffiltext{2}{NASA Exoplanet Exploration Program, Jet Propulsion Laboratory, 4800 Oak Grove Drive, Pasadena, CA 91109}
\altaffiltext{3}{Steward Observatory, 933 N. Cherry Ave, University of Arizona, Tucson, AZ 85721, USA}
\altaffiltext{4}{Konkoly Observatory, Research Center for Astronomy and Earth Sciences, P.O. Box 67, H-1525 Budapest, Hungary}
\altaffiltext{5}{Space Telescope Science Institute, 3700 San Martin Drive, Baltimore, Maryland 21218, USA}
\altaffiltext{6}{Department of Astronomy, University of Massachusetts, Amherst, MA, USA}
\altaffiltext{7}{Institute for Astronomy, ETH Z\"urich, Wolfgang-Pauli-Str. 27, 8093, Z\"urich, Switzerland}
\altaffiltext{8}{Dept of Optics and Quantum Electronics, University of Szeged, H-6720, Szeged, Hungary}
\altaffiltext{9}{NASA Ames SOFIA Science Center, N211, Mountain View, CA 94043, USA}

\email{balog@mpia-hd.mpg.de}

\begin{abstract}
We present new deep UBVRI images and high-resolution 
multi-object optical spectroscopy of the young ($\sim$ 6 - 10 Myr old), relatively nearby (800 pc) open cluster IC 2395. We identify nearly 300 cluster members and use the photometry to estimate their spectral 
types, which extend from early B to middle M.  We also present an infrared imaging survey of the central region using the IRAC and MIPS instruments on board the {\em Spitzer Space Telescope}, covering the wavelength range from 3.6 to 24\,$\micron$. 
Our infrared observations allow us to detect dust in circumstellar disks originating over a typical range of radii $\sim$\,0.1 to $\sim$\,10\,AU from the central star. We identify 18 Class II, 8 transitional disk, and 23 debris disk candidates, respectively 6.5\%, 2.9\%, and 8.3\% of the cluster members with appropriate data. We apply the same criteria for transitional disk identification to 19 other stellar clusters and associations spanning ages from $\sim$ 1 to $\sim$ 18 Myr. We find that the number of disks in the transitional phase as a fraction of the total with strong 24 $\mu$m excesses ([8] - [24] $\ge$ 1.5) increases from $8.4 \pm 1.3$\% at $\sim$ 3 Myr to $46 \pm 5$\% at $\sim$ 10 Myr. Alternative definitions of transitional disks will yield different percentages but should show the same trend.
\end{abstract}


\section{Introduction}

 The {\it Spitzer Space Telescope} \cite[{\it Spitzer};][]{wer04} has significantly improved our understanding of how protoplanetary disks form, evolve, and eventually dissipate. By 15\,Myr the accretion of gas onto protostars has largely ceased and most primordial disks have dissipated \cite[eg.][]{hai01,mam04,meng16}. By this time, planetesimals have formed and dust produced in their collisions yields planetary debris disks. These regenerated disks indirectly reveal the
presence  of  planetary bodies required to replenish the dust and allow us to trace the evolution 
of planetary systems over the full range of stellar ages \citep[e.g.,][]{lag00, dom03, wya08,gaspar13, sierchio14}. 

The beginning of the transition from an optically-thick accretion disk to an 
optically-thin debris disk occurs from the inside-out \cite[e.g.][]{skr90,sic05,meg05,muzerolle10, espaillat14} and is marked by a 
characteristic spectral energy distribution (SED) with little excess at shorter ($< 6 \mu$m) wavelengths but still retaining strong emission at the 
longer ones. These ``transitional'' disks appear to represent the process of clearing ``caught in the act''. The result is a largely evacuated inner region accompanied by an optically-thick primordial disk at larger radii. This phase is crucial to our understanding of disk dissipation and planet formation because it signals the end of stellar accretion and the consumption of nearly all the gas in the disk. Given the small number of transitional disks identified relative to the number of primordial and debris disks, it has been concluded that this phase is of short duration, on the order of a few hundred thousand years \citep{skr90,ken95,sim95,wol96, muzerolle10, espaillat14}.  

The key time period to observe the transitional phase is from a few to about 15\,Myr.
Clusters and associations are ideal laboratories for studying disk evolution as
the member stars are coeval to within a few million years, of similar composition and reddening, at similar and reliably measured distance, and numerous enough to have a wide range of masses and to support drawing statistically valid conclusions. Unfortunately, there are only a few appropriately aged young clusters within a kiloparsec to support characterizing the transitional disk phase. For more distant clusters,  {\em Spitzer} has insufficient sensitivity  in the mid-infrared to measure the photospheres of the lowest mass members and to identify complete 
samples of transitional disks. 

The open cluster IC\,2395 can augment studies of this phase of disk evolution.  The cluster is 800 pc distant \citep[Section 3.3,] []{cla03}. Despite its proximity, it has not been extensively studied. \cite{cla03} conducted the largest photometric investigation of the cluster, identifying candidate members and estimating the cluster's age, distance, extinction, and angular size. Sensitive to a limiting magnitude of $V$ $<$ 15\,mag, their survey found 78 probable and possible members through $UBV$ photometry. There have also been several proper motion studies but none combined with photometric data. The cluster age has been estmated at 6 $\pm$ 2 Myr \citep{cla03} on the traditional calibration for young cluster and association ages \citep[e.g.,][]{mam09}.  A revised age calibration has been proposed \citep[e.g.,][]{pecaut12,bell13,bell15}, on which we derive in this paper an age of $\sim$ 9 Myr  (Section 3.3). On either calibration, IC 2395 is in the critical range to characterize transitional disk behavior. 

To increase our understanding of this cluster, we have obtained $\sim45$ square arcmin fields of deep optical
($UBVRI$), and mid-IR (3.6, 4.5, 5.8, 8.0\, and 24\,$\micron$)
photometry and high-resolution optical spectroscopy of IC\,2395. 
We describe the new observations in Section 2 and discuss the cluster membership and age in Section 3. 
 In Section 4, we identify and characterize the circumstellar disks in the cluster with emphasis on 
identifying transitional disks. We combine these results with a homogeneous treatment of 
transitional disks in 19 other young clusters and associations in Section 5, to probe the 
evolution of circumstellar disks through the transitional phase. We summarize and conclude the paper in Section 6.

\section{Observations and Sample Selection}
In this section, we discuss the {\it Spitzer} observations and data reduction for  IC\,2395, as well as 
our optical UBVRI observations and spectroscopy of selected probable members. In addition to the photometry described 
below, we took JHK measurements from 2MASS. We also re-examine the results of \citet{cla03} and include appropriate members from their work in our study. 
\subsection{{\em Spitzer}/IRAC}

IC\,2395 was observed using IRAC \citep{faz04} on {\em Spitzer} on 2003 December as part of a GTO program (PID 58, PI Rieke, Evolution and Lifetimes of Protoplanetary Disks) to study protoplanetary disks and dust evolution. The survey covers a $\approx$ 44$\arcmin \times$ 44$\arcmin$ area (9 by 9 grid, $\sim$0.54 square degrees) in each of the four IRAC channels. The 12 s high-dynamic-range mode was used to obtain two frames in each position, one with 0.4 s exposure time and one with 10.4 s. The observation of each field was repeated twice with a small offset, providing 20.8 s integration time for each position. The frames were processed using the Spitzer Science Center (SSC) IRAC Pipeline v14.0, and mosaics were created from the basic calibrated data (BCD) frames using using a custom IDL package, Cluster Grinder, that treats bright source artifacts and removes cosmic ray hits and spatial scale distortion during mosaic construction \citep{gut09}. Due to the 7 arcmin offset between channels 1/3 and channels 2/4, the total area covered in all four channels is about 0.37 square degrees.

The  IDL-based photometry visualization tool PhotVis version 1.10 \citep{gut08} 
was used on the reduced images to find sources and carry out aperture photometry on them. 
The radii of the source aperture,
and of the inner and outer boundaries of the sky annulus,
were 2.4, 2.4, and 7.2 arcsec, respectively. The calibration was based on 
large-aperture measurements of standard stars. The zero point
magnitudes of the calibration were 19.6642, 18.9276, 16.8468,
and 17.3909 corresponding to zero point fluxes of 280.9, 179.7,
115.0, and 64.13 Jy for channels 1, 2, 3, and 4, respectively \citep{reach05}.
Corrections of 0.21, 0.23, 0.35, and 0.5 mag were
applied for channels 1, 2, 3, and 4, respectively, to correct for the differences
between the aperture sizes used for the IC 2395 sources and for the standard stars.

\subsection{{\em Spitzer}/MIPS}
IC\,2395 was also observed using MIPS on {\em Spitzer} in 2003 as part of the same GTO program. MIPS  is equipped with a three-channel camera with central wavelengths of approximately 24, 70, and 160\,$\micron$ \citep{rie04}. The longer wavelength channels are insensitive to stellar 
photospheric emission at the distance of IC\,2395 and no 
cluster stars were detected at 70\,$\micron$ nor 160\,$\micron$. This study is based on only 
the MIPS 24\,$\micron$ channel. 

The observations used the medium scan mode with 
half-array cross-scan offsets resulting in a total exposure time per pixel of 
80\,s.
The images were processed using the MIPS instrument team Data Analysis Tool \citep{gor05}, 
which calibrates the data, corrects distortions, and rejects cosmic rays 
during the coadding and mosaicking of individual frames. A column-dependent median subtraction routine was applied to remove any residual patterns from 
the individual images before combining them into the final 24\,$\micron$ mosaic. The total area mapped was nearly a square degree (89$\arcmin\times$40$\arcmin$).

We measured the 24\,$\micron$ flux density of individual sources using the standard photometry 
routine \texttt{allstar} in the \textsc{iraf} data reduction package \texttt{daophot}, and within a 15$\arcsec$ aperture. We then applied an aperture correction of 1.73 to account for the flux density outside the aperture, as determined from the STinyTim 24\,$\micron$ PSF model \citep{eng07}. Finally, 
fluxes were converted into magnitudes referenced to the Vega spectrum (with the zero point at 7.17 Jy). 
Typical 1-$\sigma$ measurement uncertainties for the MIPS 24\,$\micron$ fluxes are 50\,$\mu$Jy; there is also a $\sim$\,2\% uncertainty in the absolute calibration \citep{eng07}. The MIPS image is sufficiently sensitive to detect the photospheres of $\sim$\,A0 stars at the distance of IC\,2395.

\begin{figure}
\includegraphics[angle=+0,width=15cm]{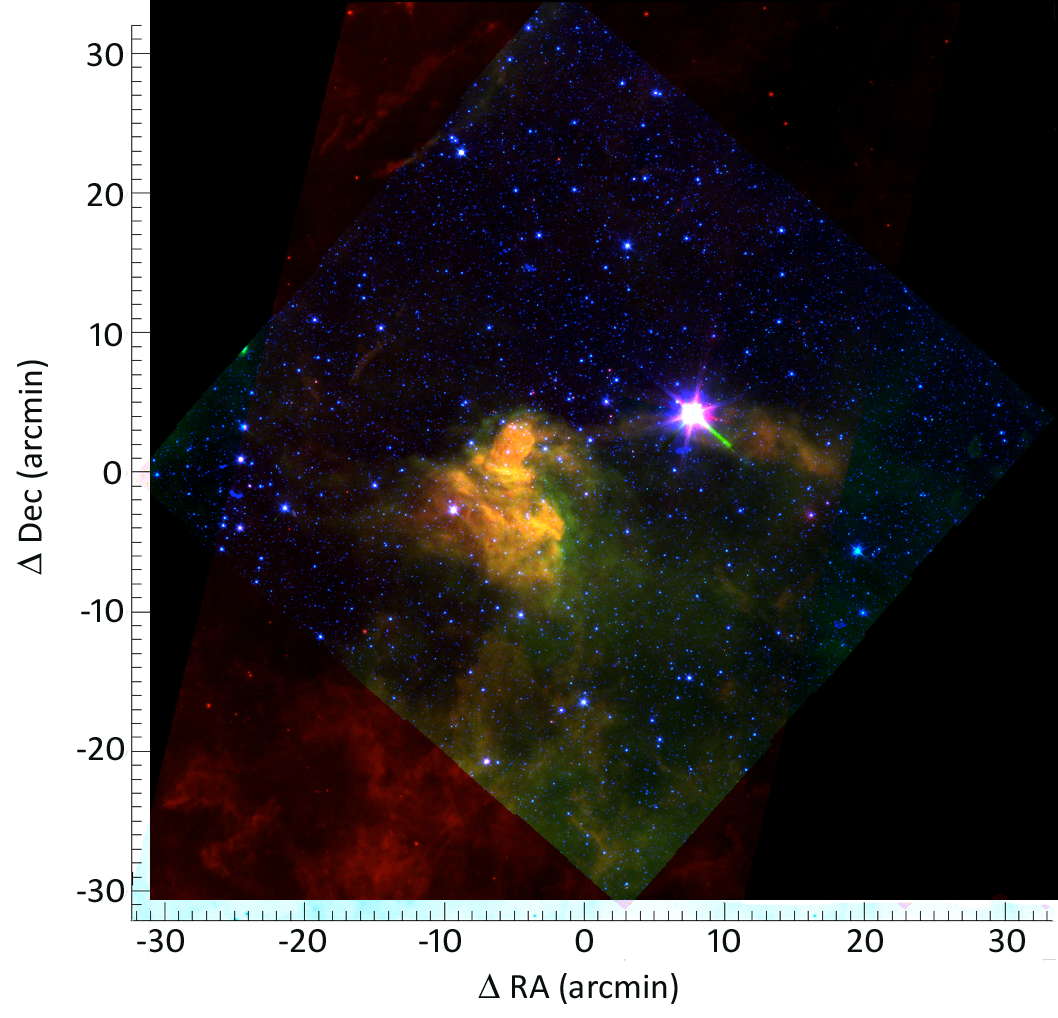}
\caption{IC\,2395 three color composite image composed of IRAC wavelengths 3.6\,$\micron$ (blue) and 8.0\,$\micron$ (green) and MIPS 24\,$\micron$ (red). The coordinates are relative to RA = 130.6111, DEC = -48.1690. Likely contributors to the emission in the 8.0\,$\micron$ channel are silicate grains and aromatic molecules, seen as the bright green areas.
\label{rgb_image}} 
\end{figure}

\begin{figure}
\includegraphics[angle=+0,width=\columnwidth]{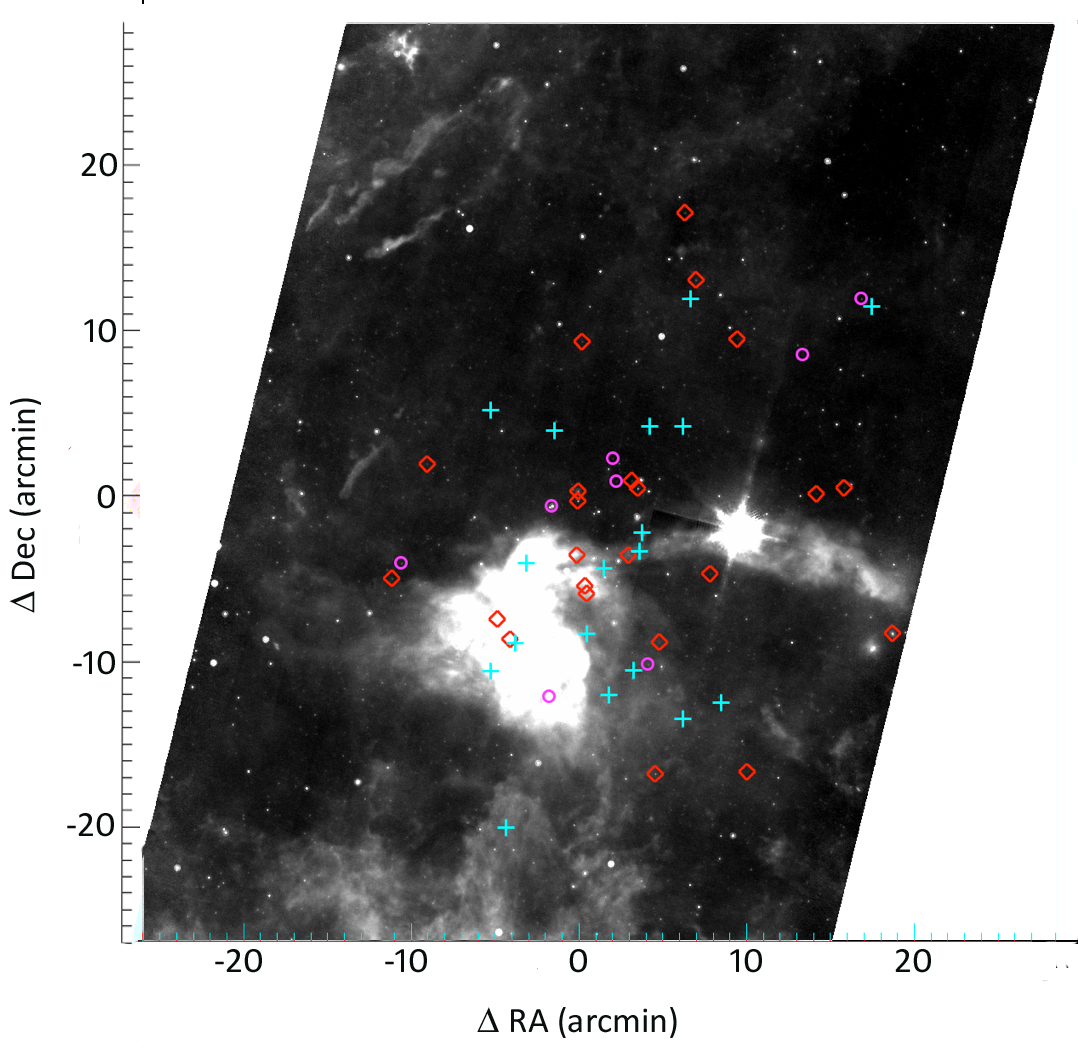}
\caption{The central $\sim$ 60$\arcmin\times$40$\arcmin$ mosaic of  IC\,2395 taken with the MIPS 24\,$\micron$ channel. The coordinates are relative to RA = 130.6389, DEC = -48.0688. Sources with disks are identified with symbols: {\it cyan crosses} represent Class II candidates, {\it magenta circles} represent transition disk candidates, and {\it red diamonds} represent debris disk candidates. The point source FWHM is 5.7$\arcsec$ and the platescale is 1.25$\arcsec$/pixel. 
\label{24um}} 
\end{figure}

Figure \ref{rgb_image} is a three-color composite image composed of IRAC wavelengths 3.6\,$\micron$ (blue) and 8\,$\micron$ (green) and MIPS 24\,$\micron$ (red). The shortest wavelength channel reaches the photospheres of all stars in the cluster. In addition, however, it also picks up many background stars. The 8\,$\micron$ image shows fewer sources and highlights dust (and associated gas). This image likely has a polycylic aromatic hydrocarbon contribution and potentially also silicate emission at its long wavelength limit. At 24 $\mu$m, we see cooler extended dust and the cluster members with significant excess emission. The brightest source is EP Velorum, an M6 asymptotic giant branch, thermally pulsating star \citep{ker94}, unassociated with the cluster. The 24\,$\micron$ mosaic of the central region of IC\,2395 is displayed in Figure \ref{24um}, with the most prominent infrared-excess sources marked. 

\subsection{UBVRI Photometry}

Our optical observations were made with the SITe 2048-\#6 CCD 
camera on the 1.5-m telescope at CTIO on 2003 Jan  23 and 24 as part of a 3 day campaign (2003 Jan  22, 23, and 24) to provide optical data in support of {\it Spitzer} observations for  IC~2395 and NGC~2451 (discussed in a separate paper; \citet{bal09}).
The  camera was mounted at the f/13.5 focal position, covering a 
$15 \times 15$ arcmin$^2$ field-of-view with a resolution of 
0.43 arcsec/pixel for the entire $2048 \times 2048$ pixel$^2$ area. 
The observations were made through Johnson-Cousins $UBVRI$ filters,
applying the Tek \#1 filter set\footnote{http://www.ctio.noao.edu/instruments/filters}. 

The whole cluster was covered by $3 \times 3 =  9$ CCD frames centered
on and around the brightest inner area at R.A. = 08:42:30, DEC =
-48:06:00. One off-cluster area (separated by $\sim 1.5$ deg from 
the cluster center) was also imaged to sample the  
foreground/background object population in the same line-of-sight. 
Each field was imaged three times through the same filter. 
One frame was obtained with a short exposure time  ($10$ s for 
$U$ and $5$ s for $BVRI$) and the other two frames were taken 
with longer ones ($250$ s for $U$, $70$ s for $B$ and $50$ s
for $VRI$). 

The reduction of the raw frames was performed with standard routines 
using {\it IRAF} \footnote{{\it IRAF} is distributed by
NOAO which is operated by the Association of Universities for
Research in Astronomy (AURA) Inc. under cooperative agreement 
with the National Science Foundation}. After trimming the edges of the
frames and subtracting the bias level from each image, 
the frames were divided by a master flat field image
obtained by median combining the available flat field frames for each filter.
Both dome flats and sky flats were taken at the beginning of each night
and combined together into the master flat frames. 
After flat field division, the two long-exposure frames corresponding to the
same filter were averaged to increase the signal-to-noise. 

The photometry of the cluster frames was conducted via PSF-fitting
using DAOPHOT implemented in {\it IRAF}. A 2nd order spatially variable PSF ({\tt varorder=2}) was built for each
frame to help compensate for the distortions of the PSFs due to
either the optical imaging artifacts in the large field-of-view, or
guiding errors that occured randomly on a few frames. The
model {\tt function=penny2} was selected to account for the slight elongation of the PSF. The
PSF-stars were selected interactively from a sample of the $\sim 100$ 
brightest, non-saturated, well isolated stars on each frame, omitting the ones 
with suspicious profiles and/or detectable neighbors within $r = 15$ pixels.
The {\tt fitrad} parameter was set according to the value of the FWHM . 
The detection threshold was fixed at the 4-$\sigma$ level on
each frame. 

The transformation of the CTIO instrumental magnitudes into the standard
Johnson-Cousins system was performed via the observations of Landolt 
photometric standard sequences \citep{lan92}. The description of the standard transformation is discussed in \citet{bal09}.

We applied aperture corrections for each frame to match the PSF photometry to the aperture photometry obtained for standard stars.  The aperture photometry was computed with 
$r_{ap} = 8$ pixels radius. The local sky level was estimated
as the mode of the pixel distribution within an annulus having inner 
and outer radii of $10$ and $20$ pixels, respectively, centered on each 
object. Inspecting the final instrumental magnitudes we found a very small ~0.02-0.03 mag systematic offset between the long and short exposure frames and also found that there is a $\sim$0.03-0.1 mag offset between the different frames of the mosaics. We tied our photometry to the middle frame (which overlaps with all of the remaining fields) of the $3 \times 3$ mosaic to ensure the consistency of our dataset.  

We tested the quality and stability of the photometry, 
including the standard transformation, 
by comparing our standard magnitudes with those from \citet{cla03} (unfortunately, only the $V$ and $B-V$ data could be compared this way). We discovered $\simeq 0.19$ mag  and $\simeq 0.015$ mag systematic offsets between the two datasets in V and B-V respectively. \citet{cla03} report that there were systematic differences (sometimes as large as 0.2 mag) between their photometry and earlier work. We therefore also compared our photometry with stars from the field found in the SIMBAD database. The systematic offsets were smaller; however, the scatter of the data was much larger due to the non-uniformity of the SIMBAD data. However, we used the exact same calibration in the case of NGC 2451 (Balog et al. 2009) where we found an almost perfect agreement with the previously published dataset of Platais et al (2001).  In cases where an object is missing from our photometric sample we adjust its \citet{cla03} photometry to match ours and give that value in the summary table for all the members.

\subsection{Spectroscopy}

We acquired AAOmega spectra using the Anglo-Australian Telescope at Siding Spring,
Australia on three nights (17-23 December 2009)  in conditions of clear skies with
1.5-2.5 arcsec seeing. 

\subsubsection{Target Selection}

To make optimal use of the 
telescope time, we pre-selected member candidates based on their positions on the color magnitude (CM) diagram and
color-color (CC) diagram. The 2MASS near-infrared photometry in particular is useful in
deselecting reddened background stars. We matched our optical photometry to the 2MASS positions 
and used the V vs.\ V-K CM diagram and the V-K vs.\ V-I CC diagram to separate possible cluster members
from the foreground and background population. We selected stars as member candidates
if their positions on the CM diagram were compatible with the cluster age, distance and reddening
allowing for errors due to binarity and age spread. First we combined the pre-main-sequence
isochrones with ages 3 and 10 Myr (traditional age calibration) of \citet{pal99} with the post-main-sequence
isochrones of \citet{mar08}. We selected these isochrones because they bracket the age
of the cluster ($\sim$ 6 Myr on this calibration) and allow some room for errors in the age estimates. Then we shifted
these isochrones to the distance modulus of IC2395 (800 pc)  and applied additional shifts to take into account the
reddening and extinction. We show the CC diagram with the selected
possible members in Figure 3. We also included all objects that showed some level of IR excess in the IRAC bands even when 
they were not covered by our optical imaging\footnote{The objects without V and I photometry (numbers 1, 2, 3, 4, 111, 134, 167, 
and 278) include one identified as a transitional disk (134), two as class II sources (111, 167), and one
as a debris disk (278): the ratio of the number of transitional to class II sources is identical to the overall value, so 
the selection of these objects does not bias our results on the incidence of transitional systems.}
Altogether 710 candidates were selected based on the above criteria. We were able
to obtain spectra of 675 of the candidates.

\subsubsection{Observations and Data Processing}

In the blue arm of the spectrograph, we used the 2500 V grating, providing $\lambda/\Delta\lambda$= 8000
spectra between 4800 \AA \  and 5150 \AA. In the red arm we used the 1700 D grating that has
been optimized for recording the Ca II IR triplet region. The red spectra range from 8350
\AA \  to 8790 \AA, with $\lambda/\Delta\lambda$ = 10000. This setup has the highest spectral resolution available
with AAOmega, suitable to measure stellar radial velocities. In total, we acquired 11 field
configurations centered on the open cluster.
The spectra were reduced using the standard Two-Degree Field data reduction pipeline.
We performed continuum normalization for the stellar spectra using the IRAF task onedspec.continuum 
and then cleaned the strongest skyline residuals using linear interpolation
of the surrounding continuum (see Balog et al. 2009 for a detailed description of the data processing
of AAOmega).

Balog et al. (2009) also describe the methodology for radial velocity determination. 
In summary, an iterative process was used to fit the atmospheric absorptions and the 
stellar radial velocity, based on synthetic stellar spectra. Our method is similar to that of the
Radial Velocity Experiment (RAVE) project (Steinmetz et al.
2006; Zwitter et al. 2008), including use of the same library of synthetic spectra. 
We required three iterations
to converge to a stable set of temperatures, surface gravities,
metallicities, and radial velocities. We estimate the velocities to be 
accurate within $\pm ~ 1 - 2$ km s$^{-1}$ for the cooler stars (T $<$ 8000 - 9000 K)  and $\pm ~ 5$  km s$^{-1}$
for the hotter ones. 

\section{Cluster Membership}
\subsection{High mass members from \citet{cla03}}
We now describe our identification of cluster members. Associating stars to stellar clusters can be challenging and is best carried out using several criteria, all of which need to be consistent with membership. The most confident membership designations are those that have photometric, kinematic, and spectroscopic measurements. In the case of IC\,2395, no previous study has 
constructed a membership list based on multiple criteria. 

\cite{cla03} conducted a $UBV$ investigation of the cluster's central  50$\arcmin\times$50$\arcmin$ region to a $V$ band limiting magnitude of 15. This survey is the starting point for our identification of the high-mass cluster members. They selected cluster members photometrically by examining the positions of the observed stars in $UBV$ color-magnitude and color-color diagrams with respect to the theoretical models of \cite{lej01}. Stars lying no more than 0.75\,mag above the zero-age main sequence (ZAMS) and deviating no more than 0.10\,mag from the CC main sequence locus were classified as cluster members. \cite{cla03} presented 61 sources meeting these photometric criteria. The CM and CC diagram positions of another 16 stars were somewhat ambiguous but they were retained as possible members. 

A comparison of 21 of these cluster members to a proper-motion-selected list from \cite{dia01} showed very good agreement; however, the uncertainties in the mean proper motion survey are sufficiently large to make the comparison inconclusive. Without a kinematic, spectroscopic, or near-infrared membership criterion to go along with the visible photometry, we believe that the \cite{cla03} classification is insufficient for providing a robust list of bona fide cluster members.

\subsubsection{Mean Cluster Proper Motion}

We add a kinematic criterion to the \cite{cla03} photometric membership list by selecting on proper motion.  
There are multiple estimates of the mean proper motion of IC\,2395
members, but some of the estimates are inconsistent. We re-estimated the mean
proper motion using the spectroscopically confirmed sample of 14 B-type cluster members
selected from \cite{cla03}. The NOMAD catalog from USNO
\citep{zac04} provides the best available proper motion for each
star \cite[usually from
Tycho-2 or Hipparcos;][]{hog00,bro97}. Twelve of the 14 had proper motions with the exceptions being HD\,74455 and HD\,74436. The variance-weighted mean $\mu_{\alpha} cos\delta$ and
$\mu_{\delta}$ values were calculated and the $\chi^2$ of each value
was calculated for the sample. One star, HD\,74251, was rejected
 due to contributing (by far) the majority of the $\chi^2$
for both $\mu_{\alpha}$cos$\delta$ and $\mu_{\delta}$. Further clipping, however, had
negligible effect on the final proper motion, so we calculated the mean proper
motion value for the remaining 11 B-stars as representative for the
group: $<\mu_{\alpha} cos\delta>$ = -3.9\,$\pm$\,0.4 mas\,yr$^{-1}$ and
$<\mu_{\delta}>$ = +3.0\,$\pm$\,0.4 mas\,yr$^{-1}$. As the
variance-weighted mean uncertainty ($\sim\,$0.3 mas\,yr$^{-1}$) was
close to the uncertainty in the Tycho-2 reference system proper motion
($\sim$\,0.25\,mas\,yr$^{-1}$), we conservatively added that
 term in quadrature to derive our final uncertainty estimate
($\sim$\,0.4\,mas\,yr$^{-1}$). The mean proper motion is consistent
within the errors whether we calculate it as a true median
\citep{got01}, a Chauvenet-criterion clipped mean \citep{bev92},
or an unweighted mean, so our choice of $\mu$ estimation matters
little. 

Our derived mean proper motion agrees well with most of the previously
measured values as shown in Table \ref{pm}
\citep{kha05,kha03,lok03,dia02,dia01,bau00},
but is severely at odds with the quoted values from
\cite{gul92} and \cite{dia06}. The \cite{dia06} value
is dominated by large numbers of faint UCAC2 stars and likely suffers
from a significant amount of field star contamination. As we (and most other
studies) do not agree with the mean proper motion estimated by
\cite{dia06}, we do not use their membership probabilities.
\subsubsection{Revised High Mass Cluster Membership List}

We now take the 61 probable and 16 possible cluster members from \cite{cla03} and further select those as members that meet the {\it combined} near-infrared and optical photometric and the proper motion criteria, i.e., those:
\begin{itemize}
\item lying near a dereddened isochrone on near-IR and optical CM and CC diagrams.
\item with proper motions within two sigma of our derived cluster mean, {\it and}
\item whose proper motion uncertainties are less than 5 mas/yr. 
\end{itemize}

 All of the \cite{cla03} objects were consistent with the photometric criterion. The second criterion was determined through a $\chi^{2}$ comparison to the mean cluster motion (as measured in \S\,3.1.1 and presented in Table \ref{pm}) which includes the objects' proper motion uncertainty along with an assumed intrinsic velocity dispersion of 1\,mas/yr, where 1\,mas/yr $\approx$\,0.7\,km/s \citep{bev92}. In equation form:

\begin{equation}
\chi^2 =
\left \{
\frac{\left [ \overline{(\mu_{\alpha} \cos \delta)^{\rm{cl}}}
- (\mu_{\alpha} \cos \delta)^*\right ]^2}
{(\sigma^{\rm{cl}}_{\rm{int},\mu_{\alpha} \cos \delta})^2
+ (\sigma^*_{\mu_{\alpha} \cos \delta})^2}
\right \}
+
\left \{
\frac{\left [ \overline{\mu_{\delta}^{\rm{cl}}}
- \mu_{\delta}^* \right ]^2}
{(\sigma^{\rm{cl}}_{\rm{int},\mu_{\delta}})^2
+ (\sigma^*_{\mu_{\delta}})^2}
\right \}
\end{equation}

\noindent
where the ``cl'' superscript designates the cluster, the ``int'' subscript designates intrinsic, and the asterisk superscript designates individual stars. By selecting those sources with $\chi^{2}\leq6$ and two degrees of freedom, we expect only $\approx$\,5\% of bona fide cluster members to be rejected using this criterion ($\sim\,2 ~ \sigma$).

We invoke the last criterion to reduce the chances that sources with relatively large uncertainties may unjustifiably obtain low $\chi^{2}$ values and contaminate our sample of bona fide cluster members. We selected a 5 mas/yr cutoff because it is less than the typical UCAC2 uncertainty for their faintest objects ($V\gtrsim12$). The consequence of this criterion, however, is that at the distance of IC\,2395, spectral types inferred from {\it J-H} to be roughly later than mid-F are deselected. In our effort to reduce interlopers, we have potentially removed faint cluster members reducing both the sample size and mass range of a measured disk fraction. Incorporating later spectral types will require adding other criteria such as radial velocity, spectral classification, or youth spectral features (see \S 3.2).

Our proper motion criteria retained 40 of the \cite{cla03} sample of 61 probable cluster members. We examined the six sources that had $\chi^{2}\leq6$ and uncertainties greater then 5\,mas/yr and found that half of them had very inconsistent mean proper motions and were left deselected. The other three had mean proper motions within 1 $\sigma$ of the cluster mean ($\chi^{2}\leq1$) and we reclassified them as possible cluster members. Another of the original 61 had no measured proper motion and was retained but only as a possible member. We also examined the sources that had $\chi^{2}\geq6$ and uncertainties less than 5\,mas/yr to identify any sources that were potentially penalized for having abnormally small reported uncertainties ($\leq$\,2\,mas/yr). Three were retained and reclassified as possible cluster members.

Of the original \cite{cla03} sample of 16 possible cluster members, three were proper motion selected and hence upgraded to probable members. One possible member had no measured proper motion and remains a possible member. 

After having applied the additional membership criteria, we end up with 43 probable cluster members and 14 possible members. As we  discuss below, we believe that two of the possible members are probable members, increasing the reported number to 45 cluster members and 12 possible members. In the following section, we discuss the selection of members from radial velocities measured from 
our spectra. In addition to allowing us to extend the membership list to lower masses, we apply this test to the possible massive 
members from \citet{cla03}, eliminating a total of ten probable and possible members\footnote{However, the radial velocity selection criterion is biased against close massive binaries, so some of the eliminated stars may in fact be cluster members.}. 

\subsection{Low mass member candidates from radial velocities}

%
%

\subsubsection{Radial Velocities}

The result of our radial velocity survey is shown in Figure 4. the cluster members are clearly
concentrated around 24.7 km s$^{-1}$ with $\sigma$ = 1.54 km s$^{-1}$. We accepted an 
object as a cluster member if its radial velocity is within 2.4 $\sigma$ (= 1 full width at half maximum of the distribution) of the mean radial velocity\footnote{We relaxed the radial velocity requirement if the object appeared to be a member or possible member based on proper motion data. In this case we accepted a star as
a member for the final analysis if its radial velocity was within 4.8 $\sigma$ of the mean radial velocity.}. Increasing this criterion to 3 $\sigma$ admits 17 additional sources, but nine are likely to be non-members, where we have estimated the number of non-members as the average for all velocities in the figure outside of 4 $\sigma$ from 24.7 km s$^{-1}$. Increasing the window to 4 $\sigma$ admits a total of 46 additional stars, but by the same method we estimate that 24 are likely to be non-members. Close binaries will have discrepant velocities outside our adopted criterion for 
cluster membership and will be rejected on this basis; it appears that there are relatively few such cases, or extending the radial velocity selection threshold would add more probable members. 

We also revised the high-mass membership of the proper
motion members based on radial velocities, reducing the total size of this sample to 37 members and 10 possible members. The final selection of members from \citet{cla03} is listed in Table 2, while the full membership list including those from Table 2 is provided in Tables 3 and 4. Figure \ref{cm_cc} places the members on CM and CC diagrams. Altogether we identified 250 low-mass members  that were not
included in the proper motion sample, based on our radial velocity survey. 

\subsubsection{Spectral Types}

Spectral types were assigned as available from the literature, as indicated in Table 2. 
The U-B vs.\ B-V color-color diagram shows that the reddening is uniform across the region, such that the U-B vs.\ B-V sequence is well-defined and there are no obvious signs of spread (outside of expectations from the presence of binaries) \citep{cla03}. We estimated the $E_{B-V}$ for the members with spectral types using the intrinsic main-sequence relation of Pecaut \& Mamajek (2013). Reddening this sequence according to $E_{U-B} = 0.73*E_{B-V}$, our estimate is $E_{B-V}$=0.09 mag, 
as also found by Claria et al. (2003). We therefore adopt an extinction equivalent to $E_{B-V} = 0.09$. 

Where types were unavailable, 
we used our photometry to estimate them.
For the stars earlier than $\sim$ K4 the preferred color for this purpose, $V - J$ or $V - K$, depends on the age of the cluster and whether there is, for example, active accretion. $V - K$ is the better choice for evolved field stars because of the longer wavelength baseline, whereas for very young star forming clusters $V - J$ is preferred to circumvent excess emission at $K$. IC 2395 is in between. Therefore, before selecting the bands we did a number of tests. First, we computed trend lines of $V$ vs. $V-J$ and $V-K$. The scatter was similar (in all these evaluations, we rejected outliers in similar numbers - $\sim$ 8\% - for both bands). However, when the scatter was weighted by the expected $V - J$ or $V-K$ value to create a metric for the uncertainty in stellar type that would result, the metric was a factor of 1.39 smaller for $V-K$. That is, the larger wavelength baseline resulted in a significant advantage for use of $V - K$. After identifying the Class II sources, we also tested whether $K$ or $IRAC1$ (hereafter [3.6]) could be contaminated by excess emission. To do so, we took all of the Class II sources and compared their observed $K - [3.6]$ colors with those we would expect from the spectral types we assigned them as discussed in the following paragraph and using the \citet{luh10} standard colors for young stars. There were no significant discrepancies, and the average was $-$ 0.01, that is the $K - [3.6]$ color was 0.01 bluer than expected for the standard colors.

Therefore, for types earlier than K4, we made the type estimates based on the (extinction-corrected) 
$V-K$ color compared with the tabulation in \citet{mam15}, while for K4 and later we used the $J - $[3.6] colors from \citet{luh10}. After a preliminary assignment of types, we determined empirical loci for the apparent $V$ and [3.6] magnitudes vs.\ type. We rejected any type estimates for stars that deviated from these loci by more than 1.1 magnitudes in $V$ or 0.8 magnitudes in [3.6]. That is, the types were only accepted if the stars were consistent in $V-K$ or $J - $[3.6] {\it and} had apparent magnitudes consistent with cluster membership at {\it both} $V$ and [3.6]. Of the 295 members, 80 failed the tests for a consistent classification, of which 13 are Class II sources (see below).


\subsection{Age}

We have put the age of the cluster on the revised age scale \citep[e.g.,][]{bell13}. To do so, we estimated the cluster age and distance from the Claria et al. (2003) catalog of probable luminous members, using the \citet{eks12} main-sequence interior models including the effects of rotation. Given that the photometry is in the apparent color-apparent magnitude plane, it is first necessary to transform the interior models into color-magnitude space and then also redden the model corresponding to E(B-V)=0.09 mag. To transform the interior models we used the \citet{cas04} so-called ODFnew models. To redden the model isochrones we used the standard A$_V$ = 3.1*E(B-V) relation (as appropriate for the low level of reddening, see e.g. \citet{ols75}). The most massive star, HD 74455, provides the best age diagnostic for this method. It sits in the vertical region of the model isochrones, hence the small dependencies of the estimated age on the color and reddening — combined with the small reddening of the cluster itself — will have an insignificant effect on the result. This star is a likely ellipsoidal variable \citep{morris85}, i.e. a binary, but given the verticle isochrones the single-star luminosity for one of the pair is still compatible with our assigned age.  Using main-sequence fitting, we estimated a distance modulus of $\sim$ 9.5 mag (equivalent to $\sim$ 800 pc; as also found by Claria et al.) and an age of $\sim$ 9 Myr.

This age is confirmed by the $V$ vs. $V - J$ HR (CM) diagram in Figure \ref{HR}, which is primarily based on the low-mass cluster members identified in our study. Both age determinations agree on $\sim$ 9 Myr .  As is usually the case, the revised calibration gives a significantly older age than the traditional estimate of 6 $\pm$ 2 Myr \citep{cla03}. 

IC 2395 is in an age range where absolute ages are not well-determined but relative ones are better understood \citep{sod14}. We  assign an error of 3 Myr, i.e., we take the age to be $9 \pm 3$ Myr. This error is to be understood as 
a statement that IC 2395 is likely to be similar in age to Upper Sco and Ori OB1b, nominally near 10 Myr (on the revised age scale), and neither so young as 
classic star-forming clusters such as $\rho$ Oph, NGC1333, NGC 2244, or IC 348 nor so old as Ori OB1a and LCC/UCL. On the traditional age scale, these clusters/associations are in the same relative sequence, but all at younger ages.

\section{Analysis}
The key result of this section is the identification of candidate IC\,2395 cluster members with evidence of circumstellar disks. {\em Spitzer} photometry is efficient in identifying evolutionary stages for disks around young stars \cite[e.g.][]{all04,meg04,har05,sic06,meg05,lad06,all07,wan07}. For the youngest systems, these studies have distinguished deeply embedded protostars (Class I) from accreting T-Tauri-like stars (Class II) from ``normal'' stars (Class III), based on placement on IRAC CC diagrams (e.g. [3.6]-[4.5] versus [5.8]-[8.0]). We build on this body of experience to separate Class II sources from the non- or weak-excess Class III cluster members and to identify the transitional disks caught in the process of transformation between these classes. At the age of IC 2395, second-generation debris disks are also starting to appear - these are systems where the dust is not primordial, but is generated in planetesimal collisions. 

We have matched the selected cluster members presented in Tables \ref{gen_m} and 3 to the IRAC and MIPS photometry using a 2.5$\arcsec$ radial positional threshold.  Several of the \cite{cla03} objects are outside the areas covered with IRAC and MIPS and a handful are incomplete in the IRAC detections. Altogether 277 objects out of 297 are detected in all 4 IRAC bands and 67 of
those also have counterparts at 24 $\mu$m. We use this body of photometry to identify the transitional, Class II, and strong debris disks in IC 2395. We find 18 Class II sources (6.5\% of the sources detected in all four IRAC bands), 8 transitional disks (2.9\% of the full IRAC detections) and 23 debris disk candidates (8.3\%). 

The most significant risk in using the longer wavelength {\em Spitzer} channels for identifying stars and protostars with strong infrared excess emission is contamination from thermal dust continuum from the residual natal molecular cloud, including emission from polycylic aromatic hydrocarbon molecules (PAHs, which contribute strongly at 8\,$\micron$), and confusion with background sources along the line-of-sight. After identifying the cluster members with strong emission from circumstellar disks, we discuss the extent to which contamination may influence these results.


\subsection{Identification of excess types among cluster members}
Figures \ref{cm_cc} and \ref{vk_k24} show the objects with excesses on different optical-near-infrared-mid-infrared CM and CC diagrams. The locations of the identified sources are also shown in Figure \ref{24um}. Two objects (\#6 and \#21) that have large K $-$ [24] excess are not classified because all of our classification schemes require IRAC data and these two stars are outside the area covered by IRAC. Based on their K $-$ [24] color they can be either transitional disks or class II sources. The identification of the other objects is explained in the next three subsections. Since we will use identical criteria for a sample of 19 clusters and associations as listed in Table 6, we discuss our approach in this general context.

\subsubsection{Transitional disks}
Although it is desirable to identify transitional disks through infrared spectroscopy \citep{espaillat14}, our aim is for a large sample to investigate their incidence and how it evolves with age. Since infrared spectroscopy is not available for all of this sample, we used photometry to test for photospheric-like colors in the 4 $\mu$m region and to measure the size of the excess at 24 $\mu$m, and also impose a requirement on the spectral type of the star for systems old enough that extreme debris disks might be confused with transitional ones. 

We first derive a criterion to test for photospheric-like colors. We start with a simple physical definition, obtained from \citet{sic08}: a transitional disk should have no excess out to 6 $\mu$m, but should retain a large excess at 24 $\mu$m, indicative of 
retention of much of the primordial disk in the more distant zone that dominates at this wavelength. In this regard, transitional disks can be distinguished from Class II sources, which have significant excesses already at 6 $\mu$m \citep{sic08}. The simplest way to isolate candidate transitional disks, then, is to use a color difference involving IRAC band 3 at 5.8 $\mu$m. We prefer [3.6] $-$ [5.8] because we will want to apply identical criteria to many clusters and associations, some with significant reddening, and this color difference is significantly less affected by extinction than is the case for color differences involving shorter wavelengths, such as $K$ \citep[e.g.,][]{flaherty07}. In addition, the measurements of [3.6] and [5.8] are obtained at the same time and hence are not affected by variability.  

To determine the acceptable values of [3.6] $-$ [5.8] to isolate candidate transitional disks from Class II sources, we show in Figure \ref{trans_select} the distribution of this color difference for our entire sample of stellar clusters and associations (listed in Table 6), along with a Gaussian fit to the primary peak in the distribution. There is a distinct excess over the Gaussian fit for sources with [3.6] $-$ [5.8] $<$ 0.4. We adopt this value as the limit for a candidate transitional disk since it defines a class of object that apparently does not belong to the population of typical Class II YSOs. The $K - [6\mu m]$ slope quoted as the upper limit for transitional disks by \citet{kim13} predicts this same color, while the limits by \citet{espaillat14} and \citet{muzerolle10} are more lenient, equivalent respectively to [3.6] $-$ [5.8] = 0.48 and 0.56. 

In the cases of Upper Sco, Lower Centaurus Crux (LCC), Upper Centaurus Lupus (UCL),  and TW Hya, we have identified transitional disks from WISE photometry. To determine a criterion similar to that we have adopted for IRAC photometry, we compared W1 - W2 vs. [3.6] $-$ [5.8] for more than 100 sources with measurements in both systems in Upper Sco, finding that the equivalent limit is W1 - W2 $<$ 0.43 with a nominal error of 0.01. 

We also need to define a minimum level of excess at 24 $\mu$m. \citet{muzerolle10} define``weak excess" transitional disks with a slope equivalent to [8] - [24] $\ge$ 1.5. They also define an optically thick disk with a slope roughly equivalent to [8] - [24] = 3.5.  We adopt a threshold of [8] - [24] = 1.5 and compare our results at this level with those at [8] - [24] $>$ 2.5 and  [8] - [24] $>$ 3.5\footnote{For comparison, \citet{cieza12} use a roughly similar threshold of [3.6] - [24] = 1.5 to identify candidate transitional disks.}. 
To determine similar criteria 
for measurements with WISE, we used measurements of sources in Upper Sco detected in both sets of photometry to set the equivalent threshold to be W3 - W4 $>$ 0.8 and W2 - W4 $>$ 1.7. 

A question remains of whether the {\it Spitzer} measurements should also put a limit on how strong the source infrared excess can be at IRAC4 (8 $\mu$m). We explored the implications of strong fluxes in this band using the YSO disk SED fitting tool \citep{robitaille06} and found that disks with masses and accretion rates well within the range expected for transitional disks (e.g., disk masses of $\sim 10^{-6} M_\odot$ and accretion $ < 10^{-10} M_\odot$   yr$^{-1}$) could have substantial fluxes at 8 $\mu$m. In addition, the broad IRAC 8 $\mu$m band can include silicate emission, which can be strong in transitional disks. Therefore, we imposed no requirement there. 

Finally, we imposed a spectral type criterion. The great majority of transitional disks are around stars 
of spectral type later than F \citep{muzerolle10,espaillat14}.  The evolution of protoplanetary disks is faster around 
higher mass stars \citep[e.g.,][]{kennedy09,yasui14}. Therefore, by an age of $\sim$ 10 Myr, we do not expect to find many transitional 
disks around early-type stars. To reflect this trend, we imposed the requirement 
that the stellar spectral type for ages $>$ 6 Myr had to be later than 
F (either from spectroscopy or estimated through photometry) to accept a disk as being transitional. The transitional disks 
in IC 2395 resulting from these criteria are indicated in Table \ref{Candidates} (WT for weak transitional, $[8] - [24] < 2.5$, and T for transitional). Our final selection criteria are also illustrated in Figure 8.

To test the degree of debris-disk contamination in these selections, we used the list of ten extreme debris systems in \citet{bal09} (we excluded HD 21362 since its excess is dominated by free-free emission). These sources have excesses at 24 $\mu$m by at least a factor of four (i.e., 1.5 magnitudes, the threshold for our identifying a weak transitional disk) and hence can be compared directly with our candidate transitional disks, since the latter are required to have similar excesses. We used the identical approaches with these sources, first testing with {\it Spitzer} [3.6] $-$ [5.8] photometry (from \citet{bal09,gor07, wein11}) and if that was lacking, using WISE measurements.  Two of the extreme debris systems are too red in these colors to pass our transitional disk criterion; if we also impose the spectral type criterion, {\it only} BD 20 307 would pass our selection. Only HR 4796A has [8] - [24] $>$ 2.5, but it is too early-type to pass our criteria. Therefore, the potential contamination appears to be small, particularly at the higher thresholds for 24 $\mu$m excess. If the [8] - [24] $>$ 1.5 threshold were allowing a significant number of extreme debris disks, we would expect the fraction of transitional disk candidates to decrease as the threshold was increased, but Table 6 shows that, if anything, there is a trend in the opposite direction. We conclude that, at least for the samples $\le$ 15 Myr in age, debris-disk contamination is not an issue.

\subsubsection{Class II Objects}
Class II objects are pre-main sequence stars characterized by SEDs with photospheric emission at visible wavelengths up to about 2\,$\micron$ followed by flat or gradually decreasing slopes at longer wavelengths \cite[e.g.][]{lad84,lad87}. The mid-infrared emission is believed to be due to large amounts of heated dust dispersed in an optically-thick, gas-rich primordial disk. In some cases, there may also be an ultraviolet excess component of the SED indicating radiation emitted from an accretion shock on the star's surface \citep{muz03}. Class II includes classic T Tauri stars (CTTSs) and Herbig AeBe (HAeBe) stars. Their existence within a cluster, especially in large numbers, implies stellar ages on the traditional scale of less than about 10\,Myr \cite[e.g.][]{hai01,mam05,hil05}. Stars older than this age have typically dissipated their primordial gas \cite[e.g.][]{pas06} so that any thermal dust emission is optically thin.

Our criteria for identifying transitional disks join consistently onto the criteria for photometric identification of Class II sources from 
\citet{gut09}, which are: [3.6] $-$ [5.8] $>$ 0.4 {\it and} [4.5] $-$ [8] $>$ 0.268 $\times$  ([3.6] - [5.8]) + 0.393. We also required a detection at 22 or 24 $\mu$m. These objects are designated II in Table \ref{Candidates}. 



\subsubsection{Debris Disks}\label{bozomath}

Debris disks represent the final evolutionary state of circumstellar disks. By this stage, the primordial gas has fully dissipated and impacts among the planetesimals create a collisional cascade, resulting in a dusty, gas-less disk. When heated by the central star, the dust reradiates in the mid-infrared. 
Stars in IC 2395 with [8] $-$ [24] $>$ 0.15 {\it and} spectral type of F or earlier {\it or} [8] $-$ [24] $<$ 1.5 and of later spectral type were designated debris disks (D in Table 5). If no spectral type could be assigned photometrically (and none was available spectroscopically), we designate the status in Table \ref{Candidates} as D?. Sources 8 and 15 are two marginal cases with [8] $-$ [24] = 0.15 at significance levels of 5 and 3.4 $\sigma$, respectively. We have not included them in Table \ref{Candidates}, but their parameters can be recovered from Tables 3 and 4 if desired. We identified 23 debris disk candidates. A number of the debris disk candidates have 
modest (0.15 - 0.30 magnitudes) IRAC band excesses at 8 $\mu$m.

\subsection{Contamination}\label{bozomath}

The most likely contaminant in our identification of stars with 24 $\mu$m excesses 
is confusion from random line-of-sight positional overlap with distant optically-faint but 
infrared-bright galaxies, planetary nebulae, and AGN. The effects of these contaminations are usually small \citep{meg04,gut08}. For example, with $\sim$2000 extra-galactic 24 $\mu$m sources per 
square degree at 0.5\,mJy \citep{pap04}, for a flux less than our completeness limit but greater 
than our detection limit, the probability of a chance 
background source observed within our matching radius of 2.5$\arcsec$ of any single  cluster member is 0.3\% [$\pi$(2.5$\arcsec^{2}$/(3600$^{2}$))$\times$2000] which means that in our sample of almost 300 stars the probability of one chance alignment is close to 100\%, but more than a few cases is unlikely. We examined the angular offsets between the 2MASS and the IRAC and MIPS positions of our candidate IR excess stars to  evaluate whether their excesses can be attributed to chance alignment. We found that the average offset between the 2MASS and IRAC coordinates is about 0.3$\arcsec \pm$0.3$\arcsec$ with a maximum separation of 0.68$\arcsec$. That is,  the probability of a chance alignment 
with a 2MASS source is around 1\%. For the objects with MIPS photometry, we found that the average  offset between the 2MASS and MIPS coordinates is 0.5$\arcsec \pm$ 0.5$\arcsec$ with a large portion of the error coming from a few sources where the distance  is larger than 1$\arcsec$. These are member candidate Nos. 37, 219, 44, 82, 198, 242, 230, and 122, with offsets respectively of 2.25$\arcsec$, 1.58$\arcsec$, 1.49$\arcsec$, 1.38$\arcsec$, 1.11$\arcsec$, 1.10$\arcsec$, 1.03$\arcsec$, and 1.02$\arcsec$  (without these sources the average error is 0.4$\arcsec \pm$ 0.3$\arcsec$ with 
no significant offset). Given the errors, an offset for a single source of 1$\arcsec$ is plausible while an offset of 1.5$\arcsec$ is unlikely but still possible; any larger offset is probably a chance alignment. Therefore, we  removed star No. 37 from the sample of infrared excess sources (although its identification as a cluster member is still valid). We visually examined the images of Nos. 219, 44, and 82 and found that No. 82 is close to a bright star in an area of relatively large background confusion, also indicated by its large measurement errors; it also was 
removed from the infrared excess sample. No. 219 has no obvious background contamination and shows a large excess even in the IRAC bands which makes it unlikely to be a spurious detection. We are left with No. 44 as the only ambiguous candidate. However, this star is on well-behaved background, so we have accepted it. To summarize, we found  two IR excess candidates (Nos. 37 and 82) that are probably contaminated, and we removed them in the final analysis.

\section{Discussion: Place of Transitional Disks in Disk Evolution}
IC 2395 includes a large proportion of transitional disks. Transitional disks were first identified from IRAS data as a class with little or no excess emission at wavelengths short of 10 $\mu$m, indicating little dust close to the star, but with strong excesses at longer wavelengths indicative of an optically thick outer disk \citep{str89}. Multiple interpretations have been advanced for this behavior \citep{wil11}. Previous studies have invoked the presence of a planet \citep{ric03,qui04}, dust evolution \citep{wil05}, or photoevaporation \citep{hol00,ale06,got06} to explain the observed ``holes". 

With the end of the {\it Spitzer} and WISE cryogenic missions, and publication of nearly all the cluster observations obtained with them, the pace of finding major new samples of transitional disks will slow substantially, making an update of their behavior timely. The most recent work along the same lines \citep{espaillat14} utilized Upper Sco as its oldest sample, but since then questions have been raised about the age of the low-mass members of this moving group \citep{her15} that may undermine it as a probe of 11-Myr-old transitional disks. IC 2395 (and other stellar clusters/associations of similar age) can provide a critical independent estimate. In the following four paragraphs, we set the scene for this analysis by discussing age scales, and then sample selection. In the next section we show that, in general, the proportion of transitional disks does not change dramatically from one cluster/association to another of similar age. This conclusion justifies our averaging of the transitional disk proportions in the following section, to derive high-weight proportions of transitional disks in young and middle-aged clusters/associations (1 - 3.5 and 4 - 6.5 Myr respectively on the tradional calibration, 1 - 6 and 8 - 13 Myr on the revised one).  

We will focus on the young stellar clusters and associations listed in Table 6, for which we show ages on both the revised and the traditional calibrations. Because ages on the traditional calibration have been determined in a variety of ways, in obtaining a homogeneous set we have given priority to those selected in the review by Eric Mamajek \citep{mam09}. Where ages on the revised calibration are not available directly, we have assigned them according to similarity in published HR diagrams for clusters/associations with directly determined ages. These cases are indicated in the notes to the table. As a test of this procedure, we evaluated Ori OB1b, which according to the HR diagram from \citet{her08}, should have an age similar to those of $\lambda$ Ori, NGC 2362, and $\gamma$ Vel, all of which are indicated to be 10 - 12 Myr on the revised scale. We took the measurements of low-mass members of Ori OB1b from \citet{bric05}, adjusted them to the distance of IC 2395, and applied extinction corrections to the individual stars according to the estimates of this parameter in \citet{bric05}. We then superimposed the resulting V vs. V-J diagram on the one for IC 2395 shown in Figure 5. The agreement is excellent indicating an age of $\sim$ 9 Myr for Ori OB1b, although there is more scatter for the stars in it, presumably because of the complications introduced by the larger and variable extinction. This agreement supports the age assignment on the revised calibration.

Importantly for this study, both the traditional and revised age calibrations place the cluster/associations into age groupings identically, as shown in Table 6. Therefore, our conclusions about the behavior of transitional disks will only change in timescale with a change in age calibration, but otherwise will remain identical.  

A critical issue is how, in each case, to define a sample of cluster members for study. 
The most conservative approach is to require that the members 
be identified through techniques {\it not} involving the mid-infrared (to avoid any possibility of introducing a bias into the infrared characteristics we use to identify transitional disks).  We have applied 
the same criteria for transitional and Class II disks as discussed above for IC 2395. Table 6 lists the ratios of transitional to (transitional + Class II) disks for each stellar cluster or association. It also shows the totals for three age ranges: $1 - 6$ Myr, $8 - 13$ Myr, and $14 - 17$ Myr (all on the revised calibration, corresponding to 1 - 3.5, 4 - 6.5, and 7 - 16 Myr on the traditional calibration). For the young clusters within 1 kpc, the totals extend down to moderately late type, low mass stars, while for the two more distant young clusters (NGC 2244 and 2362), the infrared measurements extend down only to intermediate mass stars. 
Therefore, we also show the totals omitting the two distant clusters. These values are in boldface because we believe them to 
be the most reliable indicators of transitional disk behavior. We show similar boldface numbers for the intermediate age range, 
but both groupings in the oldest age range are closer than 1 kpc.

A less-conservative approach uses the {\it Spitzer} data to identify cluster members, yielding significantly larger numbers in the samples. The results with this approach are also tabulated for the youngest age range\footnote{We do not show the values for the individual clusters.}; for the intermediate and old ranges, the cluster members are all identified without Spitzer data so this case is not shown. The results for all three levels of conservatism are identical within the errors, indicating that they are relatively robust to the member-identification approach.

\subsection{Does the Proportion of Transitional Disks Vary from Cluster to Cluster?}

Table 6 shows that, in general, the individual clusters/associations in the young ($\le$ 6 Myr on the revised scale) and middle-aged (8 - 13 Myr) categories have incidences of transitional 
disks consistent within the errors with the averages for each group (in particular, Upper Sco seems to be in line with other 
clusters and associations of similar age). This is also true for the old category, but in that case with 
very minimal counts. To make this comparison quantitative, we use the difference from the average in units of the quoted errors for all the 
young and middle-aged systems. The result is a nearly perfect normal distribution, except that the number of transitional disks 
in $\rho$ Oph is low at 2/123 or 1.6\%. Turning to the infrared-selected sample, this cluster is still low in transitional disks, 9/249 or 3.6\%, but now  at less than 3$\sigma$ from the average. That is, there is no compelling evidence for variations in the fraction of transitional disks  over the full set of results reported in Table 6, although the case of $\rho$ Oph deserves further investigation. 

\citet{sic08} have reported that the Coronet Cluster may have an abnormally high incidence of transitional disks, and indeed the counts
reported for it in Table 6 are higher than average, albeit not at a statistically significant level. To understand any possible differences, 
we examine the seven transitional disks identified by \citet{sic08} in more detail. Only two of these disks are within the selection criteria 
for our sample - CrA-4111 and G-14. A third source on their list, G-65, has no data at 24 $\mu$m (Table 3 of \citet{sic08}) and furthermore 
has a [3.6] $-$ [5.8] color of 1.01, far larger than our selection threshold of $<$ 0.4. Two more sources from their list, CrA-466 and G-87, also have
large values of [3.6] $-$ [5.8], 0.81 and 0.72 respectively. Two more from their list, CrA-205 and CrA-4109, are missing IRAC data used in our 
photometric selection; extrapolating from the existing data, they would likely be classified as transitional disks if full IRAC measurements 
were available. We do not add them to our sample because a similar detailed examination has not been performed for all the other clusters. 
However, we have included one source, G-30, which they excluded because of low signal to noise at 24 $\mu$m (the indicated SNR of 6.7 puts it 
above our WT threshold by $\sim$ 1$\sigma$; excluding it would introduce a potential bias in our method). If we added the two sources with missing IRAC data without subtracting the one with a low SNR, it would not raise the apparent excess of transitional disks to a 
statistically significant level. It is necessary to add {\it three} additional disks to the three we have identified 
to barely reach a 2-$\sigma$ result. We conclude that the Coronet cluster does not 
provide a convincing counter-example to our conclusion that the incidence of transitional disks is {\it not} variable for stellar 
groupings of similar age, within the limits of the existing data.

\subsection{Change of Transitional Disk Proportion with Age}

The results of the preceding section suggest that we can use the average incidences of transitional disks for young, middle-aged, and elderly systems without significant loss of information. There is a substantial and highly statistically significant increase in the fraction of transitional disks 
among those disks with strong 24 $\mu$m excesses, going from the young to the middle-aged clusters/associations; the elderly clusters/associations have an incidence 
perhaps similar to that for the middle-aged ones, but the low numbers in the elderly grouping make any firm conclusions impossible. Figure 10 shows this trend graphically. The trend of 
an increasing incidence of transitional disks with age has been found previously, e.g., \citet{muzerolle10}, but with marginal statistical 
significance for the older clusters/associations, and by \citet{currie11}, again with sparse representation of older clusters/associations, and with the results summarized in 
the review by \citet{espaillat14}. This trend is put on a firm statistical basis by the average results for the young and middle-aged stellar clusters/assocations 
summarized in Table 6. In general, other studies using different criteria for identifying transitional disks will find them in different numbers (see 
cautions in \citet{espaillat14}), but our study remains valid in terms of trends because it uses a homogeneous selection throughout.

What are the implications of the substantial change in transitional disks as a fraction of strong-infrared-excess disks? There are two 
limiting possibilities: 1.) that there is a systematic change in disk properties with age that is reflected by an increasing probability of any given 
disk entering this stage; or 2.) that the disks retain similar structures but there is an increasing fraction of those remaining at any time that enters the transitional stage. An indication that the second case is closer to correct is that the relative fractions of disks 
with [8] $-$ [24] $>$ 1.5, 2.5, or 3.5 is virtually identical for 
the averages for the young and middle-aged stellar groupings,
 i.e., $1$: $0.98 \pm 0.21$: $1.78 \pm 0.42$ and $1$: $0.93 \pm 0.34$: $1.35 \pm 0.28$, respectively (based on the boldfaced values 
in Table 6). Since the excess at 
24 $\mu$m is an indicator for the optical depth of the disks outside the inner few AU, these statistics indicate that a minority of 
systems can retain disks that are still very dense in this zone for 10 Myr or a bit longer. 

We therefore assume that the time to clear an optically thick disk is independent of the age of the stellar grouping to which it belongs, and 
that such clearing passes through a transitional disk stage, which is of similar duration for all disks \citep[e.g.,][]{muzerolle10, espaillat14}.
The consequence of the increase in transitional disk incidence is then that the decay of the 24-$\mu$m-dominant optically thick disk component of a YSO population 
cannot be exponential, unlike that at shorter infrared wavelengths \citep[e.g.,][]{ribas14}, but must start slowly and accelerate 
relative to an exponential.

\section{Conclusions}

The open cluster IC 2395 can add significantly to our understanding of protoplanetary and early debris disk evolution, since it is 
relatively close (800 pc) and at a critical age where protoplanetary disks are disappearing and debris disks begin to dominate.
However, the cluster has largely been overlooked in disk studies. We report optical and infrared photometry and high resolution optical spectroscopy of the cluster, from which we: 
\begin{itemize}
\item Increase the list of probable members to nearly 300, spanning spectral types of early B to middle M;
\item Estimate an age of $9 \pm 3$ Myr on the revised age scale, e.g. that of \citet{bell13}; this value compares with $6 \pm 2$ Myr on the traditional scale \citep{cla03}; and
\item Identify 18 Class II (6.5\% of the members with full IRAC data), 8 transitional disk (2.9\%), and 23 debris disk candidates (8.3\%).
\end{itemize}

We have combined the transitional disk information with homogeneously defined similar objects in nineteen additional young
clusters and associations to quantify the evolution of this phase; finding that 
\begin{itemize}

\item The dominant cause of variations in the proportion of transitional disks is age; most clusters of similar age have similar proportions of transitional disks among the systems with strong 24 $\mu$m excesses. The single possible exception is $\rho$ Oph, where transtional disks are relatively rare. 

\item The relative numbers of disks with different degrees of 24 $\mu$m excess do not change significantly with age, implying that the change in the proportion of transitional disks is not driven by a systematic change of disk properties, e.g., a thinning of disks that makes them more susceptible to dissipation

\item  The number of disks in the transitional phase as a fraction of the total with strong 24 $\mu$m excesses ([8] - [24] $\ge$ 1.5) increases from $8.4 \pm 1.3$\% at $\sim$ 3 Myr to $46 \pm 5$\% at $\sim$ 10 Myr; alternative definitions of 
transitional disks will yield different percentages but should show the same trend.

\item Under the conventional assumption that the lifetime of the transitional stage is fixed, and given the evidence that the nature of the individual Class II and transitional disks does not change with age, this result implies that the decay in the proportion of systems with strong 24 $\mu$m excesses cannot be exponential, but must start more slowly and finish more rapidly than the ``best fit'' exponential. 

\end{itemize}

We have also demonstrated that IC 2395 is a rich cluster at a critical age for circumstellar disk evolution, worthy of 
additional study.

\section{Acknowledgements}

We thank Eric Mamajek for assistance on estimating cluster ages and Lynne Hillenbrand for a short course on the intricacies of current age determination. We also thank the anonymous referee for a detailed critique that yielded significant improvements in the paper. Partial support for this work was provided by NASA through Contract Number 1255094 
issued by JPL/Caltech. LLK has been supported by the Lend\"ulet Young Researchers Program of the
Hungarian Academy of Sciences. This research has made use of the SIMBAD database, 
operated at CDS, Strasbourg, France. This publication makes use of data products from the Two Micron All Sky Survey, which is a joint project of the University of Massachusetts and the Infrared Processing and Analysis Center/California Institute of Technology, funded by the National Aeronautics and Space Administration and the National Science Foundation.
It also makes use of data products from the Wide-field Infrared Survey Explorer, which is a joint project of the University of California, Los Angeles, and the Jet Propulsion Laboratory/California Institute of Technology, funded by the National Aeronautics and Space Administration. We thank the Anglo-Australian Observatory and Cerro Tololo Inter-American Observatory for granting telescope time and for logistical support of our program.


\bibliographystyle{apj}
\bibliography{ic2395}

\begin{deluxetable}{lcccccl}
\tabletypesize{\tiny}
\tablecaption{Proper Motion Studies of IC\,2395 \label{pm} }
\tablewidth{0pt}
\tablehead{
\colhead{} &
\colhead {$\mu_{\alpha}$cos$\delta$ } &
\colhead{Uncertainty} &
\colhead {$\mu_{\delta}$} &
\colhead{Uncertainty} &
\colhead{Sample} &
\colhead{} \\
\colhead{Survey} &
\colhead {mas/yr} &
\colhead {mas/yr} &
\colhead {mas/yr} &
\colhead {mas/yr} &
\colhead{size} &
\colhead {Comments}}

\startdata
\citet{gul92} & +2.3 & 1.0 & -1.8 & 1.0 & 8 & Greater then 3$\sigma$ away from this paper's result.\\
\citet{bau00} & -3.6 & 0.6 &  2.2 & 0.5 & 1 &   \\
\citet{dia01} & -4.4 & 1.6 &  4.0 & 1.6 & 11&   \\
\citet{dia02} & -4.37& 1.73&  4.05& 1.73& 14&   \\
\citet{lok03} & -4.86& 0.33&  3.48& 0.23& 33&   \\
\citet{kha03} & -4.21& 0.63&  1.73& 0.63& 12&   \\
\citet{kha04} & -4.34& 0.38&  2.42& 0.35& 15 &   \\
\citet{dia06} & -5.56& 0.40&  6.02& 0.40& 55& Greater then 3$\sigma$ away from this paper's result. \\
\hline\hline
This Paper & -3.92& 0.33&  3.03& 0.26& 11& 1 clipped; proper motion values used in this analysis. \\
This Paper & -3.60& 0.36&  2.69& 0.28& 10& 2 clipped \\
\enddata
\end{deluxetable}


\begin{deluxetable}{rcccccrrrrl}
\tabletypesize{\tiny}
\tablecaption{General Characteristics of IC\,2395 Cluster Members from \citet{cla03} \label{gen_m}}
\tablewidth{0pt}
\tablehead{
\colhead{ID\tablenotemark{a}} &
\colhead{ID} &
\colhead{} &
\colhead{} &
\colhead{} &
\colhead{} &
\colhead{{\it V}} &
\colhead{{\it B-V}} &
\colhead {$\mu_{\alpha}$cos$\delta$ } &
\colhead {$\mu_{\delta}$} &
\colhead {Other} \\
\colhead{} &
\colhead{Tbl 4\tablenotemark{b}} &
\colhead{RA (2000)} &
\colhead{Dec (2000)} &
\colhead{SpT} &
\colhead{Ref.}  &
\colhead{(mag)} &
\colhead{(mag)} &
\colhead {(mas/yr)} &
\colhead {(mas/yr)} &
\colhead {Designation\tablenotemark{c}} 
}

\startdata
 246 &  1 &  8:40:09.9 &   -48:07:57.6 &    A5V  & 1  & 11.29 &  0.10 &     -3.4$\pm$2.9 &      4.2$\pm$3.8 &                     -     \\ 
232 & 2  &  8:40:18.9 &   -48:29:43.4 &  A7V & 1  &11.30 &  0.16 &     -8.1$\pm$1.4 &      5.7$\pm$1.4 &       CD-484010     \\ 
229 & 3  &   8:40:25.6 &   -48:22:48.1 &  F7  &  3  & 12.96 &  0.56 &    -13.7$\pm$5.8 &     12.7$\pm$5.8 &               -     \\ 
 113 &  14 & 8:40:46.8 &   -48:12:42.8 &   B8: &  2 & 9.60 &  0.05 &     -8.7$\pm$3.0 &     -1.6$\pm$3.0 &         JW8, CD-474208     \\ 
 112 &  16 &   8:40:48.2 &   -48:14:15.6 &  A4 &  3  & 11.75 &  0.17 &     -6.1$\pm$2.3 &      2.0$\pm$3.2 &         JW5, CD-474209     \\ 
 111 &  23 &  8:40:53.4 &   -48:13:31.8 &    B2IV & 2  & 6.96 & -0.16 &     -3.6$\pm$0.6 &      2.2$\pm$0.4 &           JW2, HD74234     \\ 
 260 &  -- &  8:40:59.8 &   -47:47:02.8 &  F0II  & 1  & 11.39 &  0.17 &     -4.6$\pm$2.1 &      3.9$\pm$2.0 &             CD-474211     \\ 
 118 & 36  &  8:41:04.0 &   -48:04:15.6 &   F9 & 3  & 13.31 &  0.61 &      6.6$\pm$5.1 &      5.8$\pm$5.1 &                     -     \\
\enddata
\tablenotetext{a}{From \citet{cla03}}
\tablenotetext{b}{If no cross-reference to Table 4 is provided, we do not confirm the membership assignment.}
\tablenotetext{c}{CD or CPD = photographic plates from the Annals of the Cape Observatory, L = Lynga, HD = Henry Draper}
\tablenotetext{d}{Probable binary}
\tablenotetext{e}{~Likely ellipsoidal variable \citep{morris85}, hence binary}
\tablecomments{(i) Celestial coordinates are from 2MASS; (ii) optical photometry from \citet{cla03}; uncertainties described therein; (iii) proper motions from NOMAD; \citet{zac04}; (iv) mean cluster proper motion derived in \S\,3.1.1 is $\mu_{\alpha}$cos$\delta$ = -3.9$\pm$0.4\,mas/yr and $\mu_{\delta}$ = 3.0$\pm$0.4\,mas/yr.  Spectral type references: 1, \citet{pickles10}; 2, \citet{skiff14}; 3, from photometry (see text); 4, from H$\beta$ photometry \citep{paun15} calibrated through \citep{craw66}.}
\end{deluxetable}


\begin{deluxetable}{rcccccccccccccccc}
\tabletypesize{\tiny}
\rotate
\tablecaption{Optical Photometry of Cluster Members}
\tablewidth{0pt}
\tablehead{
\colhead{} &
\colhead{RA(2000)} &
\colhead{DEC(2000)} &
\colhead{} &
\colhead{U} &
\colhead{(err)} &
\colhead{B} &
\colhead{(err)} &
\colhead{V} &
\colhead{{(err)}} &
\colhead{{R}} &
\colhead {(err)} &
\colhead {I} &
\colhead {(err)} \\
\colhead{ID} &
\colhead{(deg)} &
\colhead{(deg)} &
\colhead{SpT\tablenotemark{a}} &
\colhead{mag}  &
\colhead{(mag)} &
\colhead{(mag)} &
\colhead {(mag)} &
\colhead {(mag)} &
\colhead{(mag)} &
\colhead{(mag)} &
\colhead{(mag)} &
\colhead{(mag)} &
\colhead {(mag)} 
}
\startdata

1	&	130.04111	&	-48.13267	&	A1	&	11.463	&	--	&	11.251	&	--	&	11.114	&	--	&	--	&	--	&	--	&	--	\\
2	&	130.07856	&	-48.49538	&	A1	&	11.602	&	--	&	11.32	&	--	&	11.116	&	--	&	--	&	--	&	--	&	--	\\
3	&	130.10667	&	-48.38002	&	F7	&	13.549	&	--	&	13.381	&	--	&	12.783	&	--	&	--	&	--	&	--	&	--	\\
4	&	130.13353	&	-48.15396	&	--	&	--	&	--	&	--	&	--	&	--	&	--	&	--	&	--	&	--	&	--	\\
5	&	130.16329	&	-48.35785	&	K8	&	17.685	&	0.014	&	17.052	&	0.009	&	15.836	&	0.011	&	15.122	&	0.012	&	14.41	&	0.022	\\
6	&	130.16755	&	-47.80875	&	M7	&	--	&	--	&	--	&	--	&	19.652	&	0.036	&	18.143	&	0.016	&	16.566	&	0.029	\\
7	&	130.17279	&	-48.08578	&	M0	&	19.469	&	0.058	&	18.233	&	0.013	&	16.791	&	0.008	&	15.777	&	0.015	&	14.822	&	0.013	\\
8	&	130.17401	&	-47.97556	&	A8	&	11.835	&	0.015	&	11.591	&	0.005	&	11.285	&	0.011	&	11.098	&	0.011	&	10.903	&	0.014	\\
\enddata
\tablenotetext{a}{All spectral types in this table are estimated from photometry (see text for details)}
\tablecomments{2MASS celestial coordinates are used except for ID\,110 where IRAC coordinates are used.}
\end{deluxetable}


\begin{deluxetable}{lcccccccccccccccc}

\tabletypesize{\tiny}
\rotate
\tablecaption{Infrared Photometry of Cluster Members}
\tablewidth{0pt}
\tablehead{
\colhead{}  &
\colhead{J\tablenotemark{a}} &
\colhead{(err)} &
\colhead{H\tablenotemark{a}} &
\colhead{(err)} &
\colhead{K$_S\tablenotemark{a}$} &
\colhead{(err)} &
\colhead{[3.6]} &
\colhead{(err)} &
\colhead{([4.5]} &
\colhead{(err)} &
\colhead{[5.8]} &
\colhead{(err)} &
\colhead{[8.0]} &
\colhead{(err)} &
\colhead {[24]} &
\colhead {(err)} \\
\colhead{ID} &
\colhead{(mag)} &
\colhead{(mag)} &
\colhead{(mag)} &
\colhead{(mag)}  &
\colhead{(mag)} &
\colhead{(mag)} &
\colhead{(mag)} &
\colhead {(mag)} &
\colhead{(mag)} &
\colhead{(mag)} &
\colhead{(mag)} &
\colhead{(mag)} &
\colhead{(mag)} &
\colhead{(mag)} &
\colhead{(mag)} &
\colhead {(mag)} 
}
\startdata

1	&	10.931	&	0.024	&	10.887	&	0.021	&	10.87	&	0.021	&	--	&	--	&	10.86	&	0.004	&	--	&	--	&	10.729	&	0.012	&	--	&	--	\\
2	&	10.861	&	0.023	&	10.837	&	0.024	&	10.81	&	0.019	&	--	&	--	&	--	&	--	&	--	&	--	&	--	&	--	&	--	&	--	\\
3	&	11.812	&	0.027	&	11.466	&	0.024	&	11.414	&	0.023	&	--	&	--	&	--	&	--	&	--	&	--	&	--	&	--	&	--	&	--	\\
4	&	15.07	&	0.049	&	14.308	&	0.048	&	13.813	&	0.052	&	13.351	&	0.01	&	13.175	&	0.012	&	13.09	&	0.043	&	12.38	&	0.044	&	--	&	--	\\
5	&	13.304	&	0.024	&	12.776	&	0.029	&	12.592	&	0.026	&	--	&	--	&	12.601	&	0.009	&	--	&	--	&	12.491	&	0.044	&	--	&	--	\\
6	&	14.965	&	0.049	&	14.301	&	0.058	&	13.969	&	0.067	&	--	&	--	&	--	&	--	&	--	&	--	&	--	&	--	&	9.231	&	0.016	\\
7	&	13.679	&	0.027	&	12.978	&	0.022	&	12.699	&	0.026	&	12.695	&	0.007	&	12.656	&	0.009	&	12.624	&	0.03	&	12.603	&	0.053	&	--	&	--	\\
8	&	10.593	&	0.023	&	10.506	&	0.021	&	10.459	&	0.023	&	10.495	&	0.003	&	10.448	&	0.004	&	10.395	&	0.008	&	10.396	&	0.009	&	10.245	&	0.029	\\
\enddata
\tablenotetext{a}{JHK photometry from 2MASS}

\end{deluxetable}



\begin{deluxetable}{rcccccccl}
\tabletypesize{\tiny}
\tablecaption{Disk Candidates in IC\,2395\label{Candidates}}
\tablewidth{0pt}
\tablehead{
\colhead {} &
\colhead {} &
\colhead {[3.6]-[5.8]} &
\colhead {err} &
\colhead {[4.5]-[8]} &
\colhead {err} &
\colhead{[8]-[24]} &
\colhead{err} &
\colhead{Disk} \\
\colhead {ID} &
\colhead{SpT} &
\colhead{(mag)} &
\colhead{(mag)} &
\colhead{(mag)} &
\colhead{(mag)} &
\colhead{(mag)} &
\colhead{(mag)} &
\colhead{Type} 
} 

\startdata
14	&	--	&	-0.014	&	0.005	&	0.162	&	0.006	&	1.154	&	0.034	&	D?	\\
25	&	M6	&	0.467	&	0.050	&	0.756	&	0.055	&	2.795	&	0.057	&	II	\\
29	&	M6	&	0.36	&	0.062	&	0.81	&	0.070	&	3.873	&	0.069	&	T	\\
36	&	F9	&	0.103	&	0.022	&	0.142	&	0.028	&	1.685	&	0.036	&	D?	\\
41	&	F5	&	0.137	&	0.016	&	0.199	&	0.020	&	0.829	&	0.042	&	D	\\
44	&	--	&	0.096	&	0.040	&	0.099	&	0.058	&	1.694	&	0.092	&	WT	\\
63	&	G2	&	0.197	&	0.014	&	0.022	&	0.016	&	0.428	&	0.051	&	D	\\
71	&	--	&	0.122	&	0.029	&	0.205	&	0.047	&	1.416	&	0.066	&	D?	\\
\enddata
\end{deluxetable}

\begin{deluxetable}{lcccccccccccc}
\tabletypesize{\tiny}
\rotate
\tablecaption{Proportion of Transitional Disk Candidates Among Known Members of Clusters and Associations}
\tablewidth{0pt}
\tablehead{
\colhead{Cluster or} &
\colhead {Revised Age } &
\colhead{Ref.} &
\colhead {Trad. Age } &
\colhead{Ref.} &
\colhead{Ref.}  &
\colhead {Trans./Total} &
\colhead{\%} &
\colhead{Trans/Total}   &
\colhead{\%}   &
\colhead{Trans/Total}   &
\colhead{\%}   &
\colhead{Ref}   \\
\colhead{Association}   &
\colhead{(Myr)}  &
\colhead{age}  &
\colhead{(Myr)}  &
\colhead{age}  &
\colhead{Mshp}  &
\colhead{[8]-[24]$>$1.5}  &
\colhead{}  &
\colhead{[8]-[24]$>$2.5}  &
\colhead{}  &
\colhead{[8]-[24]$>$3.5}  &
\colhead{}  &
\colhead{photometry} \\
\colhead{(1)}   &
\colhead{(2)}  &
\colhead{(3)}  &
\colhead{(4)}  &
\colhead{(5)}  &
\colhead{(6)}  &
\colhead{(7)}  &
\colhead{(8)}  &
\colhead{(9)}  &
\colhead{(10)} &
\colhead{(11)}
}
\startdata
Coronet  &   --  &  --  & 0.5 - 1  &  15 & 15  &  3/13  & 14 $\pm$ 10   &  2/8   & 25 $\pm$ 18  & 1/2 & 50 $\pm$ ... &  15 \\
$\rho$ Oph  & 0.5 - 3 & 1  & 1 - 2  & 16 & 35  & 2/123  & 1.6 $\pm$ 1.2  & 2/90 & 2.2 $\pm$ 1.6  & 0/24 & 0 $\pm$ ... & 47 \\
NGC 2068/70  &  --  &  --  & $\sim$ 2 & 17 & 17  &  7/51  &   14 $\pm$ 5  &  7/40  &  17 $\pm$ 7  & 5/7 &  71 $\pm$ 32 & 17 \\
NGC 2244 & $\sim$ 2 & 2 & $\sim$ 2 &  18 &  36  &  15/95  &  16 $\pm$ 4 &  13/78 & 17 $\pm$ 5 & 7/31 & 23 $\pm$ 9 & 48 \\
Serpens  &  0.5 - 3  & 3  & 1 - 3 & 19 & 37  & 6/79 &  7.6 $\pm$ 3.1  &  3/67  & 4.5 $\pm$ 2.6   &  2/28  &  7 $\pm$ ... & 49 \\
Taurus   &  3 - 4  &  4  &  1.5 & 20 & 38 &  10/152 &  6.6  $\pm$ 2.1  &  8/129   & 6.2  $\pm$ 2.2  & 6/54 & 11  $\pm$ 5 & 38 \\
NGC 1333 & $\lesssim$ 6 &  5 & 1 & 20, 21 & 39  &  2/44  & 4.5 $\pm$ 3.2  &  1/31 & 3.2 $\pm$ ...  & 1/9 & 11 $\pm$ ... & 47 \\
Chameleon I  &  $\sim$ 6  &  6  & 2.6 & 23 &  40 &  4/27  &  15 $\pm$ 8  &  3/20  & 15 $\pm$ 9   & 2/5 & 40 $\pm$ ... & 23 \\
\enddata
\end{deluxetable}

\clearpage

\clearpage

\clearpage
\begin{figure}
\includegraphics[angle=0,width=16cm]{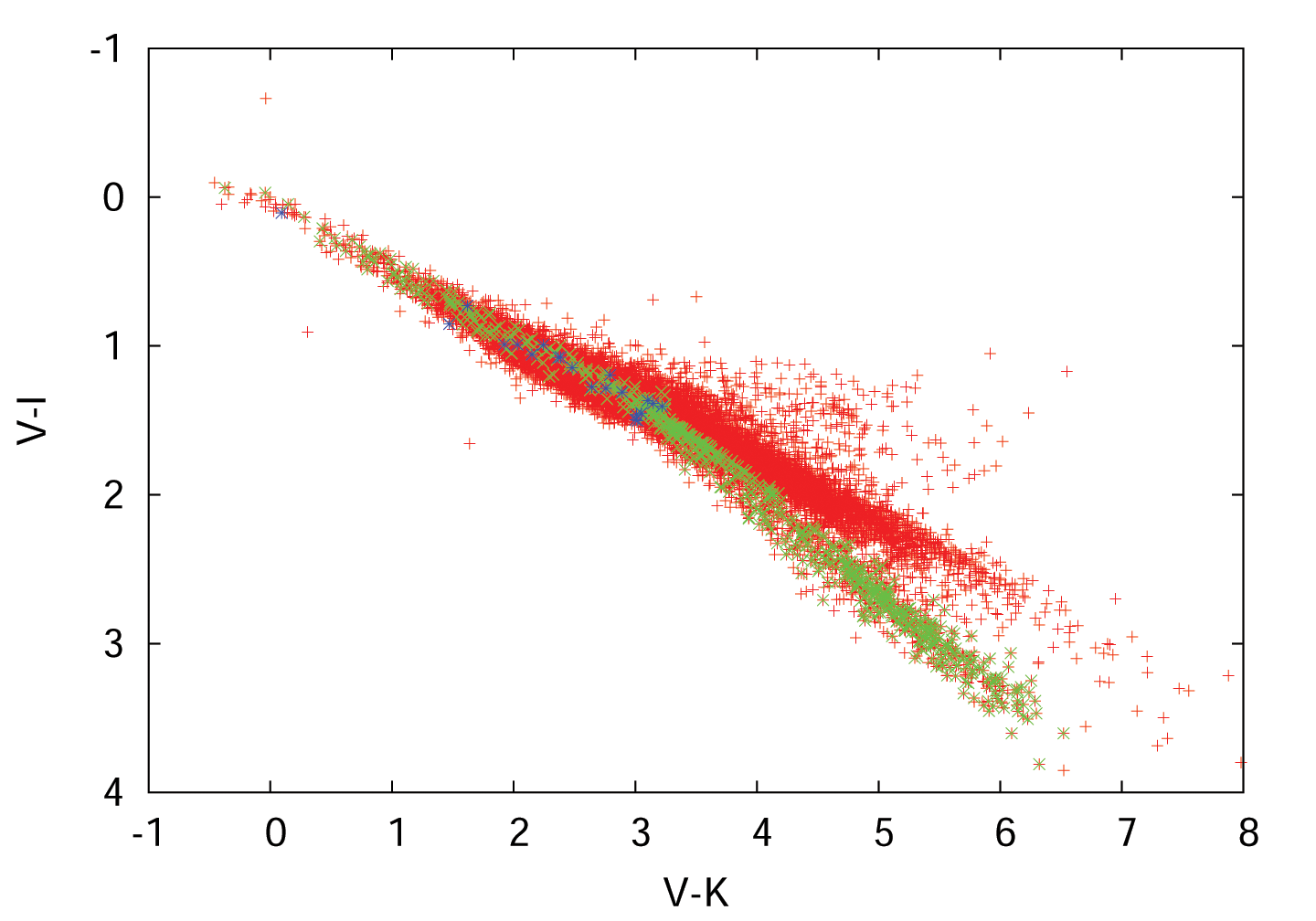}
\caption{Optical - near-infrared CC diagram of IC\,2395 cluster member candidates (green symbols) selected for the spectroscopic survey together with the complete sample of the photometric data (red symbols). Blue symbols show the stars with IR excess that may not be cluster members but were included in the spectroscopic sample.
\label{cm_cc_forspec}} 
\end{figure}

\begin{figure}
\includegraphics[angle=+0,width=5.0in]{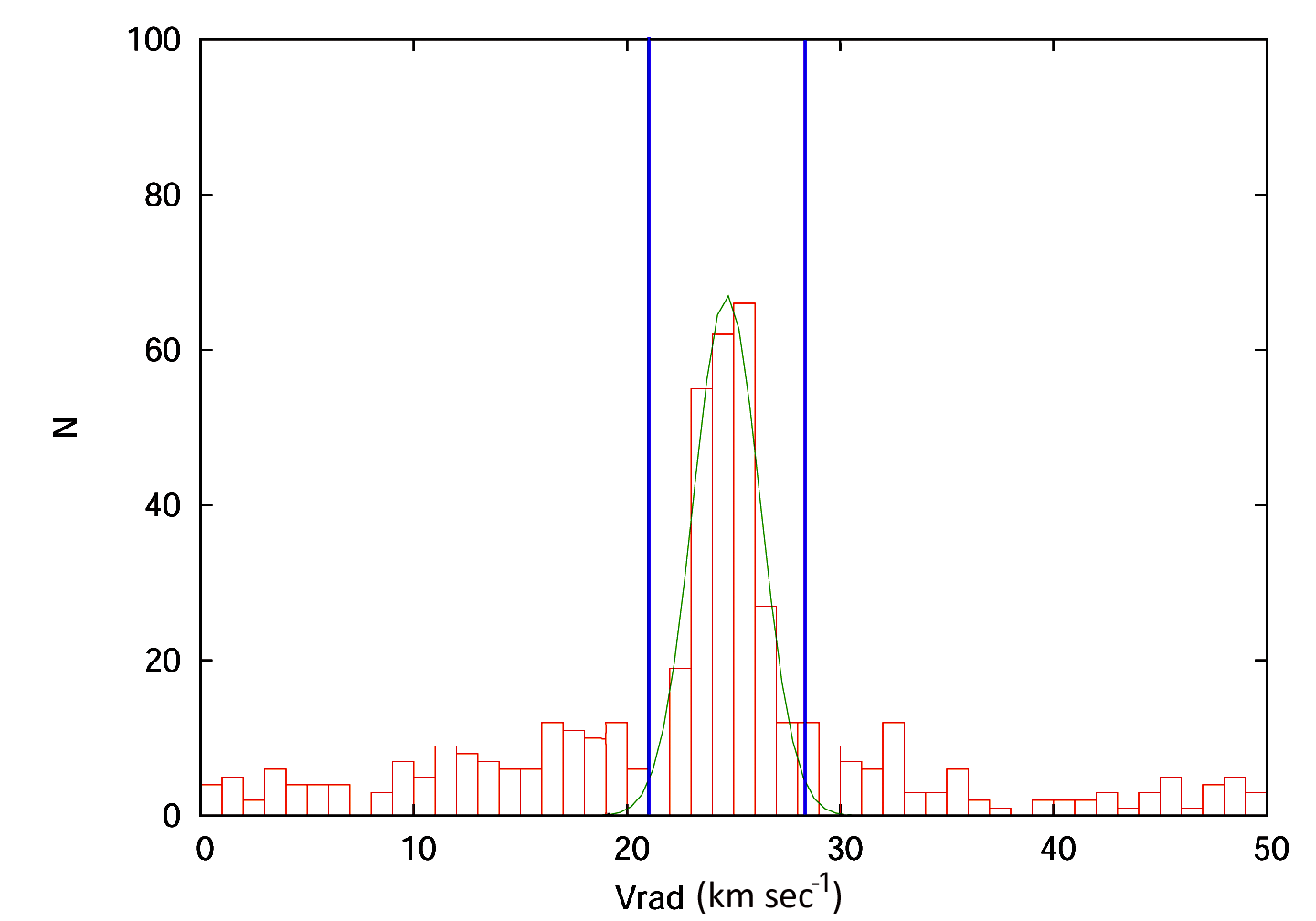}
\caption{Radial velocity distributions toward IC 2395. The Gaussian fit is centered at 24.7 km/s. The vertical blue lines indicate $\pm$ one full width at half maximum above and below the center velocity. This velocity range was used to identify cluster members; it corresponds to $\pm$ 2.4 $\sigma$. }
\label{color-color} 
\end{figure}

\begin{figure}
\includegraphics[angle=+0,width=5.0in]{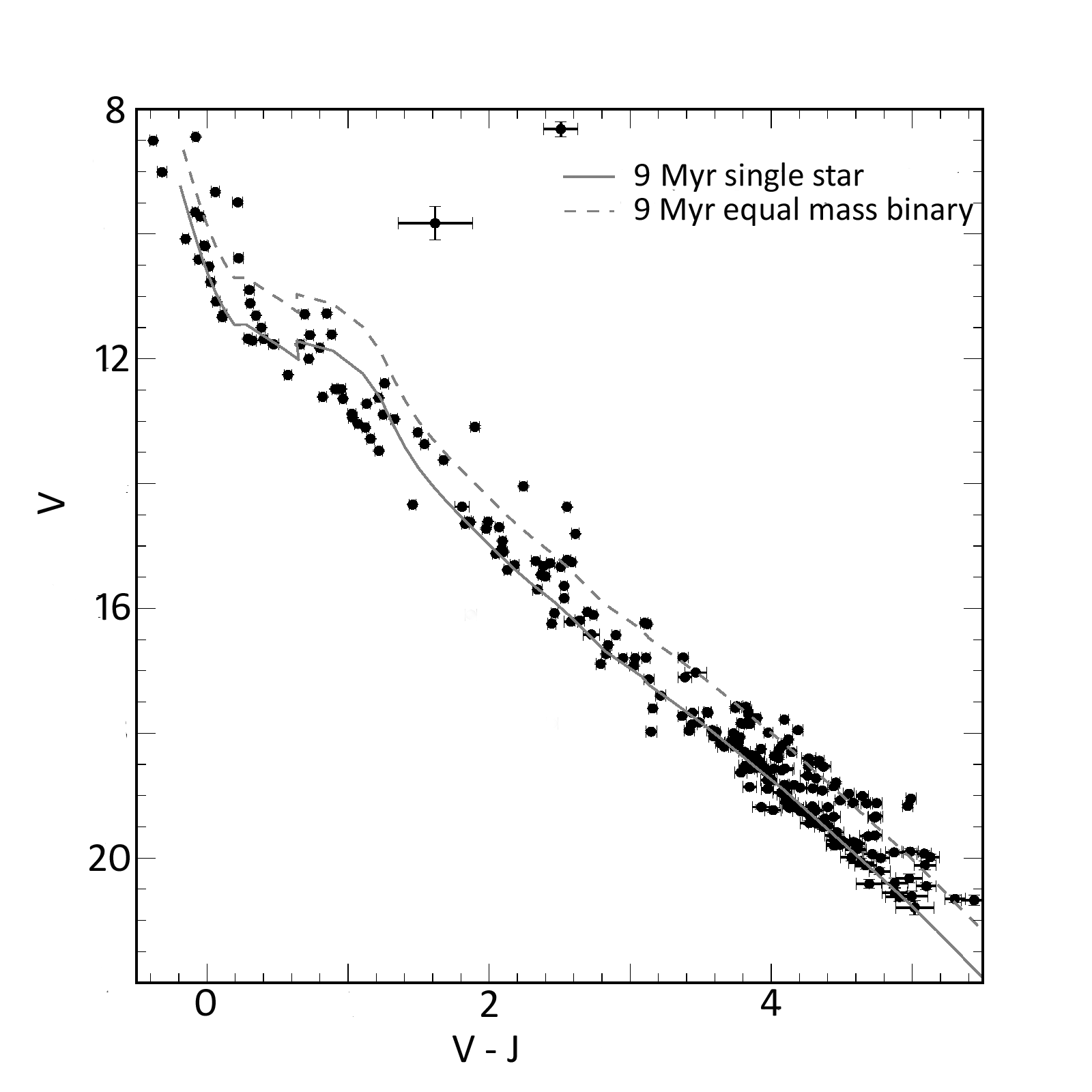}
\caption{V, V-J CM diagram with semi-empirical (see \citet{bell14} for details) model isochrones using the Dartmouth interior models.The red solid isochrone is for 9 Myr; it has been reddened by E(B-V)=0.09 and shifted in distance assuming a modulus of 9.5  The dashed line is the same but shifted 0.75 mag higher i.e. where we would expect the equal mass binaries to lie.}
\label{HR} 
\end{figure}

\clearpage
\begin{figure}
\includegraphics[angle=0,width=12cm]{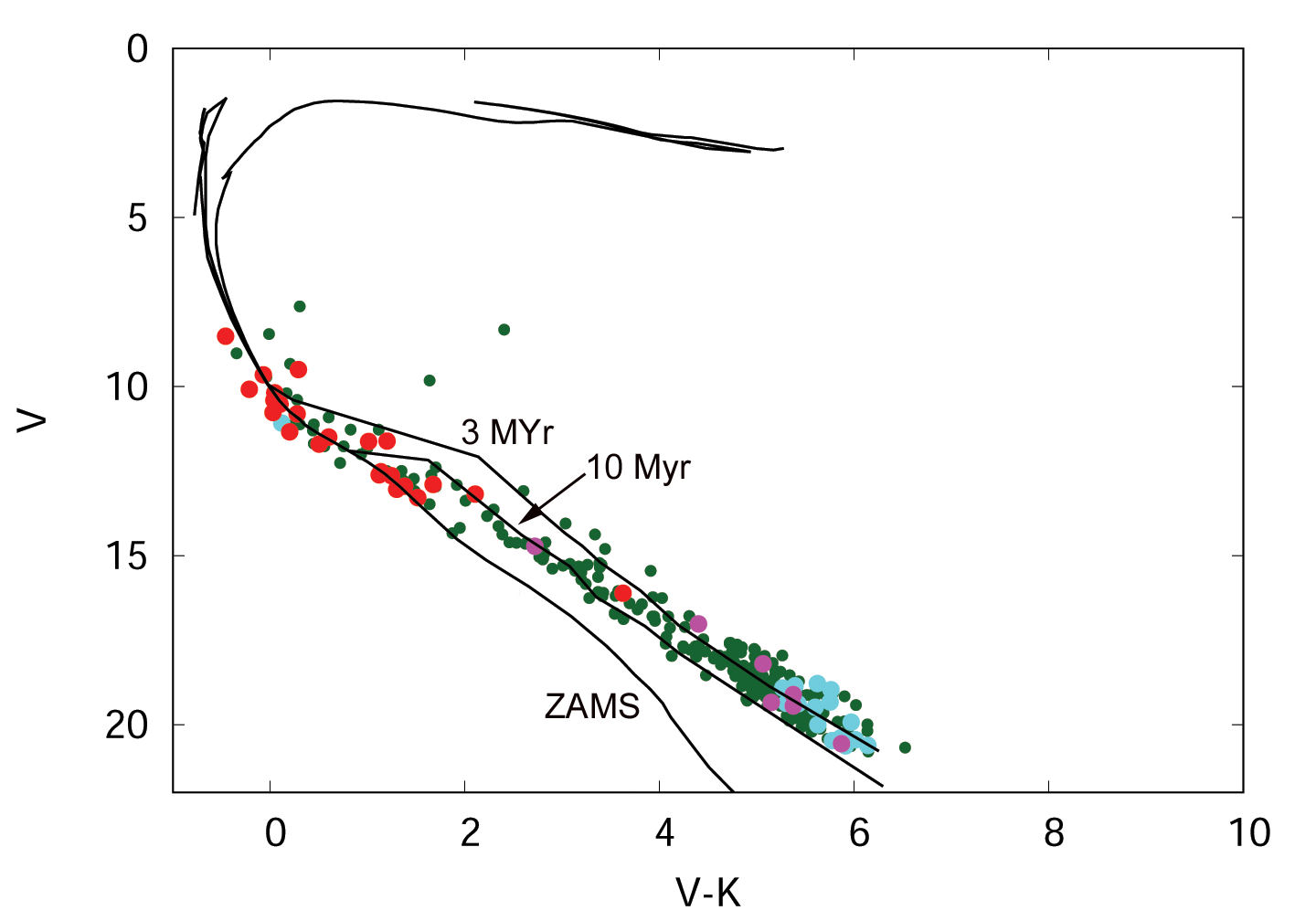}
\includegraphics[angle=0,width=12cm]{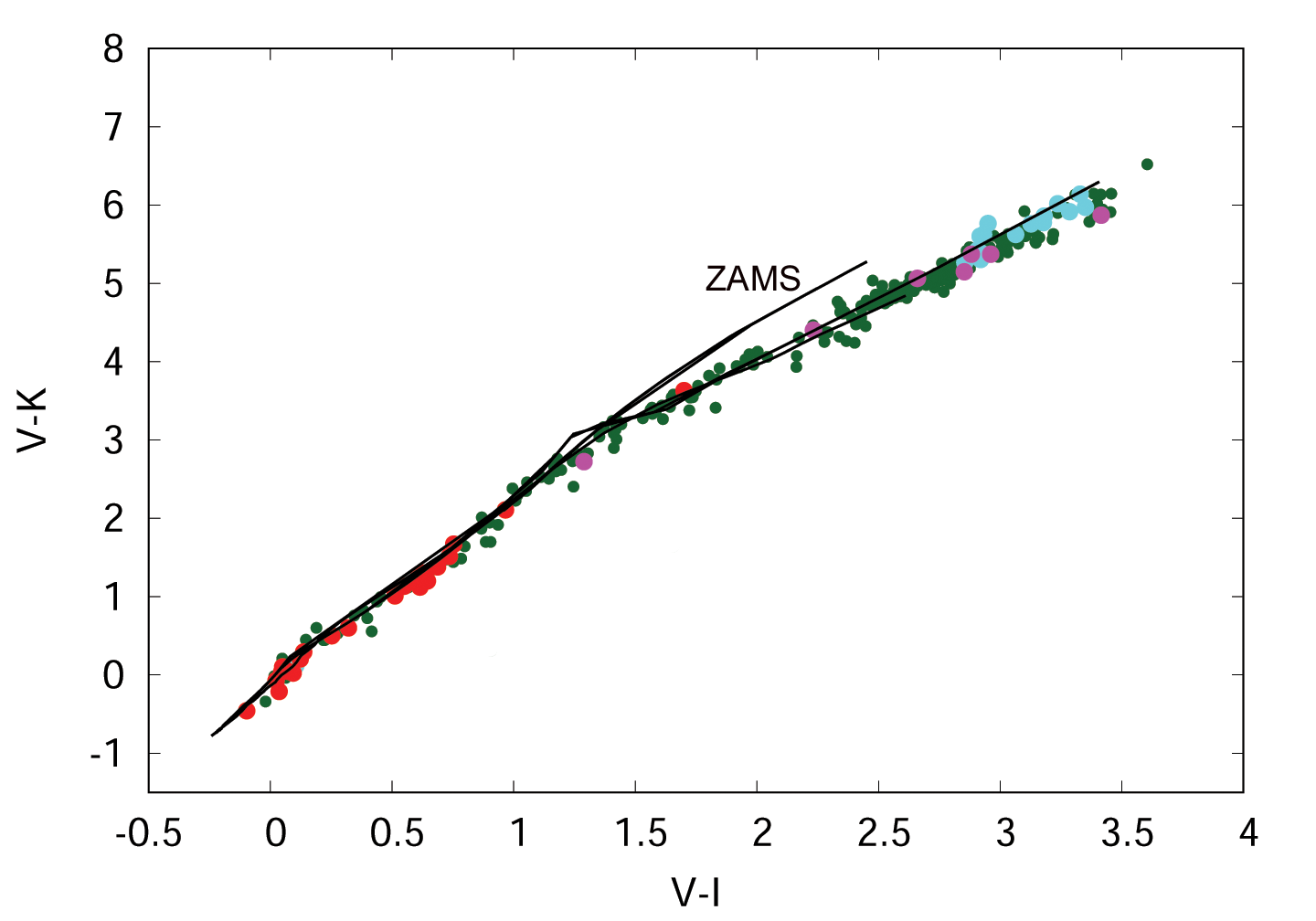}
\caption{Optical - near-infrared CM and CC diagrams of IC\,2395 cluster members. Green dots: members without IR excess, red: debris disk candidates,  magenta: transitional disk candidates, cyan: class II candidates. Black lines show the 10- and 3Myr isochrones \citep{pal99} (traditional age scale) and the ZAMS \citep{mar08}.
\label{cm_cc}} 
\end{figure}

\clearpage
\begin{figure}
\includegraphics[width=6in]{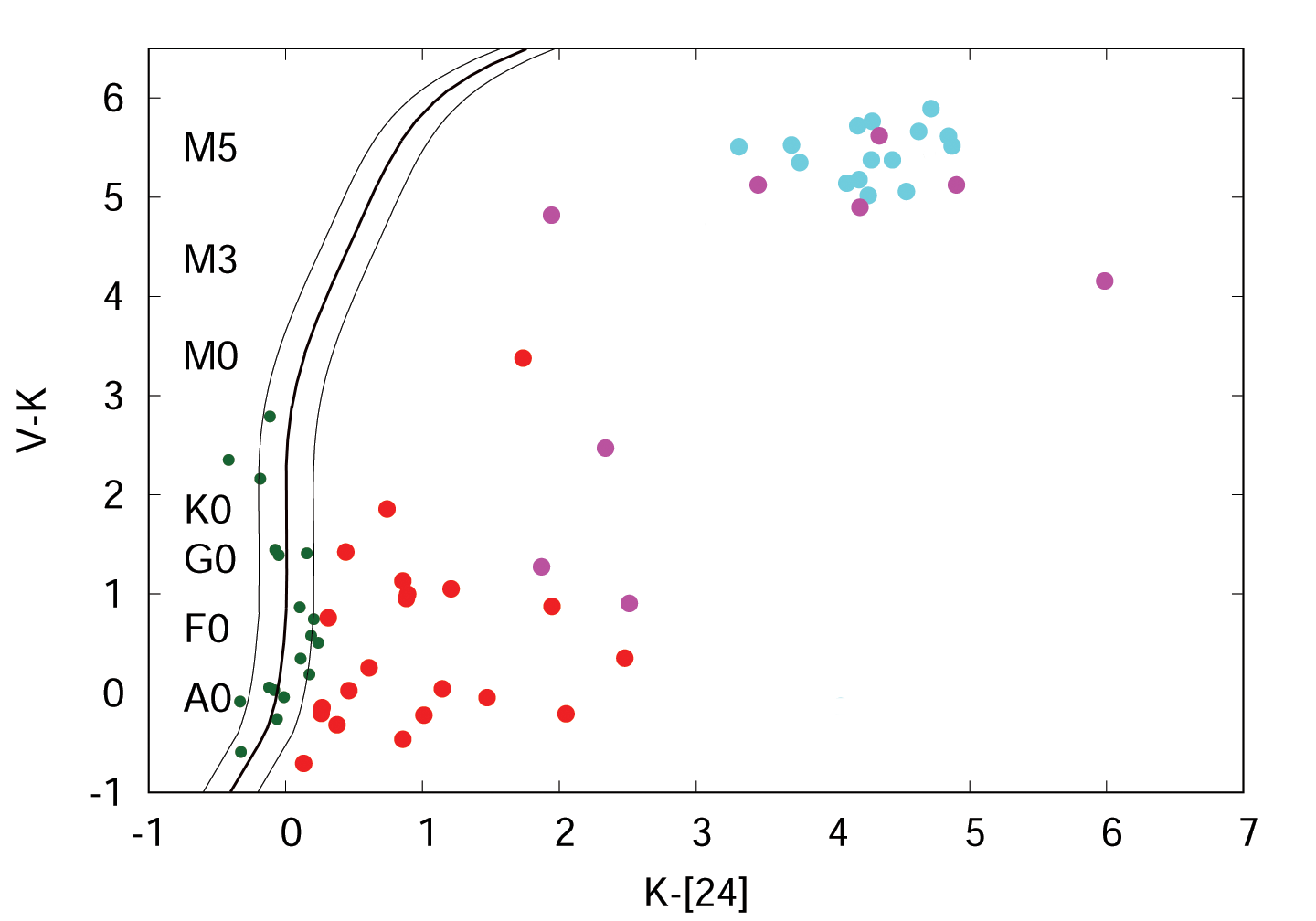}
\caption{Dereddened {\it V-K} versus {\it K$_s$}-[24.0] CC diagram. Symbols are as in Figure \ref{cm_cc}. The black line represent the photospheric colours of main sequence stars \citep{urb12}, while the bounding lines are the typical range of uncertainties 
in projecting the color from shorter wavelengths (including random and systematic errors).
\label{vk_k24}} 
\end{figure}

\clearpage
\begin{figure}
\includegraphics[angle=00,width=5in]{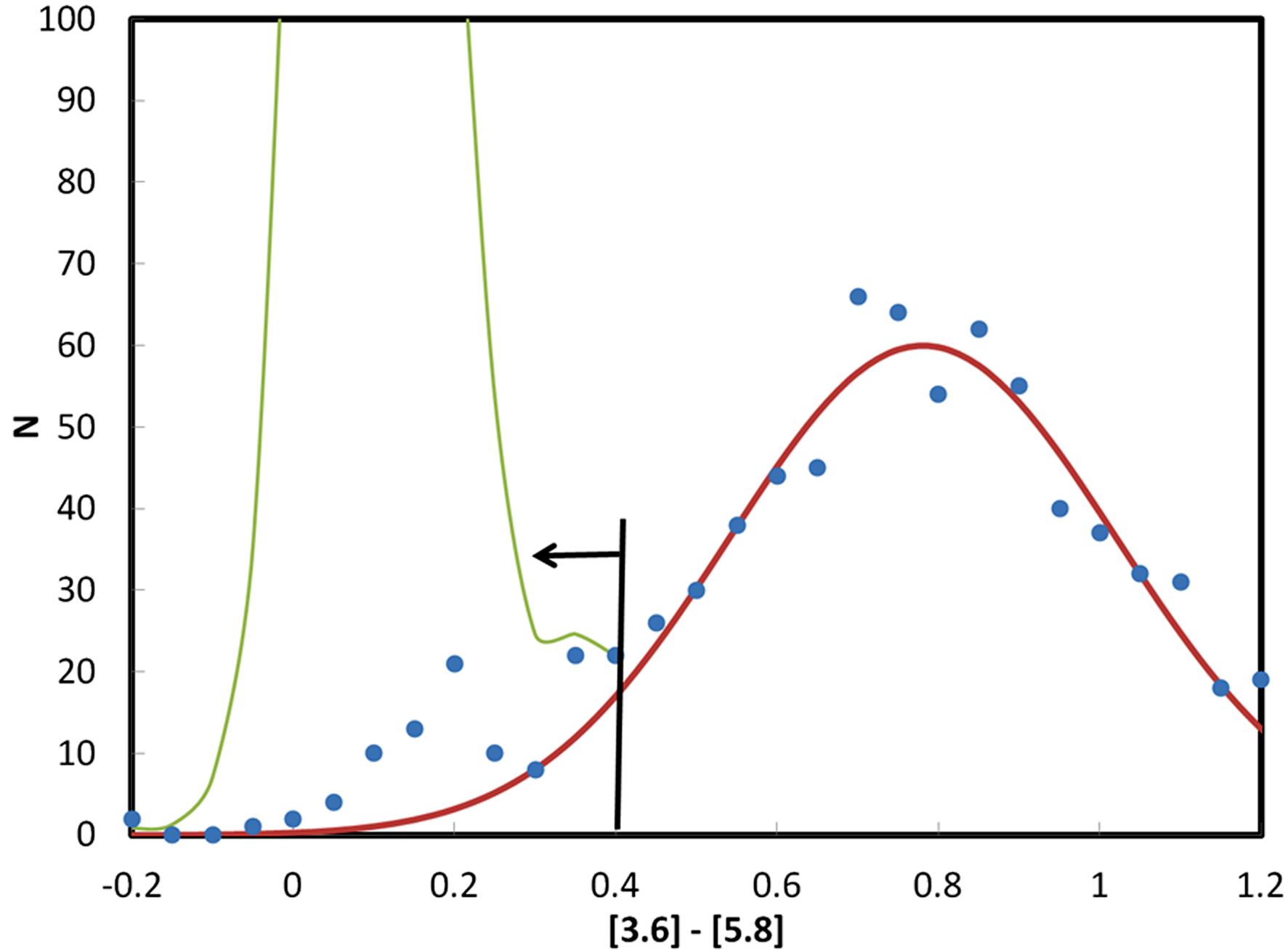}
\caption{Selection criterion for transition disks using IRAC colors. The points show the distribution of the colors for all the objects with [8] $-$ [24] $>$ 1.5 in all the clusters under study (see text), to which we have fitted a Gaussian (solid red line) confined to the upper 2/3 of the points to avoid 
biasing the fit with the wings of the distribution. The thin green line shows the distribution for the sources with smaller [24] excess emission.  The vertical line shows the criterion adopted to identify transitional disk candidates (i.e., [3.6] $-$ [5.8] $<$ 0.4).} 
\label{trans_select}
\end{figure}

\clearpage
\begin{figure}
\includegraphics[angle=00,width=6in]{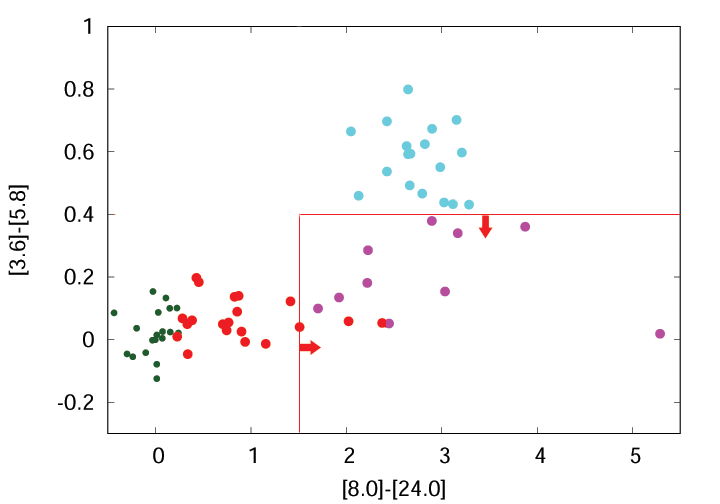}
\caption{Color-color diagram illustrating the selection criteria {\bf (within the red box)} for transitional disks (i.e., [3.6] $-$ [5.8] $<$ 0.4 \& [8] $-$ [24] $>$ 1.5) applied to IC 2395. Symbols are as in Figures 6 and 7.
} 
\label{CC_select}
\end{figure}

\clearpage
\begin{figure}
\includegraphics[angle=00,width=6in]{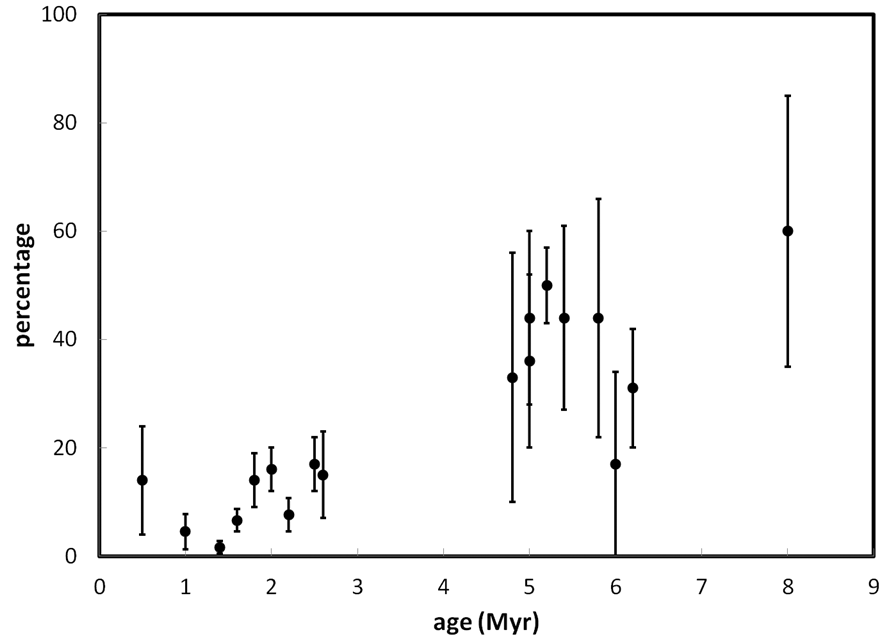}
\caption{Percentage of transitional disks vs. age. defined as No. Transitional/(No. Transitional + No. Class II) * 100. This graph uses the traditional age scale because it is more complete for the relevant clusters/associations.
} 
\label{CC_select}
\end{figure}

\end{document}